\documentclass[%
reprint,
superscriptaddress,
amsmath,amssymb,
prr,
floatfix
]{revtex4-2}

\usepackage{graphicx}
\usepackage[hidelinks]{hyperref}
\usepackage{cleveref}
\usepackage{tikz}
\usetikzlibrary{tikzmark,decorations.pathreplacing,calligraphy}
\usepackage{mathtools}
\usepackage[normalem]{ulem}

\definecolor{forestgreen}{RGB}{34,139,34}
\definecolor{burntorange}{RGB}{204, 85, 0}

\begin{document}

\title{Towards nonlinear thermohydrodynamic simulations via the Onsager-Regularized Lattice Boltzmann Method}

\author{Anirudh Jonnalagadda}
\email[Corresponding author: ]{anirudh.jonnalagadda@iit.it}
\affiliation{Computational mOdelling of NanosCalE and bioPhysical sysTems (CONCEPT) Lab, Istituto Italiano di Tecnologia, 16152 Genova, Italy}
\affiliation{Center for Life Nano- \& Neuro-Science, Fondazione Istituto Italiano di Tecnologia, Viale Regina Elena 295, Rome, 00161, Italy}
\author{Amit Agrawal}
\affiliation{Department of Mechanical Engineering, Indian Institute of Technology Bombay, Mumbai 400076, India}
\author{Atul Sharma}
\affiliation{Department of Mechanical Engineering, Indian Institute of Technology Bombay, Mumbai 400076, India}
\author{Walter Rocchia}
\affiliation{Computational mOdelling of NanosCalE and bioPhysical sysTems (CONCEPT) Lab, Istituto Italiano di Tecnologia, 16152 Genova, Italy}
\author{Sauro Succi}
\affiliation{Istituto per le Applicazioni del Calcolo, Consiglio Nazionale delle Ricerche, Via dei Taurini 19, Rome, 00185, Italy}
\affiliation{Center for Life Nano- \& Neuro-Science, Fondazione Istituto Italiano di Tecnologia, Viale Regina Elena 295, Rome, 00161, Italy}
\affiliation{Department of Physics, Harvard University, 17 Oxford St., Cambridge, 02138, MA, USA}
\date{\today}

\begin{abstract}
\noindent
This work presents a generalized, assumption-free, and stencil-independent theoretical analyses of the recently proposed Onsager-Regularized (OReg) lattice Boltzmann (LB) method [Jonnalagadda \textit{et al.}, \href{https://doi.org/10.1103/PhysRevE.104.015313}{\textcolor{blue}{Phys. Rev. E 104, 015313 (2021)}}] and demonstrates its ability to mitigate spurious errors associated with the insufficient isotropy of standard first-neighbor lattices without the inclusion of any external correction terms.
The hydrodynamic limit recovered by the OReg scheme is derived for two equilibrium distribution functions, namely the so-called thermal guided equilibrium and the popular second order polynomial equilibrium, to show that the OReg scheme yields macroscopic dynamics that are $\mathcal{O}(u)$ times more accurate than that of the bare BGK collision model.
Specifically, we show that, with the guided equilibrium on the D2Q9 standard lattice, the OReg scheme inherently compensates for the insufficient lattice isotropy of the standard D2Q9 lattice by automatically adjusting the lattice viscosity, yielding $\mathcal{O}(u^4)$ and $\mathcal{O}(u^2)$ accurate kinetic models when operated at the reference and arbitrary temperatures respectively.
Further, we also show that the OReg scheme presents an $\mathcal{O}(u^3)$ accurate kinetic model for the D2Q9 lattice when used with the second order polynomial equilibrium formulation.
Thereafter, the accuracy of the OReg-guided-equilibrium kinetic model is numerically demonstrated for quasi-one-dimensional simulations of the rotated decaying shear wave and isothermal shocktube problems.
The present work lays the theoretical foundation of a generic framework which can enable fully local, correction-free, nonlinear thermohydrodynamic LB simulations on standard lattices, thereby facilitating scalable simulations of physically challenging fluid flows.
\end{abstract}


\maketitle

\section{Introduction}

\vspace{-0.5em}
The evolution of a Lattice Boltzmann (LB) system \citep{benzi1992review, chen-doolen-ARFM, aidun2010ARFM, tiribocciPhysReps2025, succi2001lattice, kruger, succi2018lattice} emulates the Boltzmann transport equation in $D$-dimensions through microscopic operations on discrete populations, $f_i$, in a lattice-discretized velocity space having $N$ velocities $c_{i_\alpha}$, $i \in \left\lbrace 1, \cdots, N \right\rbrace$.
In the single relaxation framework, the lattice-BGK update is defined as:

\vspace{-2em}
\begin{align}
	\label{eq:lattice-update-std}
	f_i(x_\alpha + c_{i_\alpha}\Delta t,& t + \Delta t) - f_i(x_\alpha, t) = -\frac{1}{\tau}f_i^{(neq)}(x_\alpha, t) \nonumber\\
			&= -\frac{1}{\tau}\left(f_i(x_\alpha, t) - f_i^{(eq)}(x_\alpha, t)\right),
\end{align}
\vspace{-0.5em}

\noindent
where, $\tau$ is the relaxation time, and $f_i^{(eq)}$, $f_i^{(neq)} = \Big(f_i - f_i^{(eq)}\Big)$ are the equilibrium and non-equilibrium contributions of $f_i$ respectively.
In such a discrete framework, exactly recovering Navier-Stokes-Fourier (NSF) macrodynamics places explicit constraints on both the equilibrium and non-equilibrium contributions of $f_i$.

\begin{figure}[t]
    \centering
    {
        \includegraphics[scale=1]{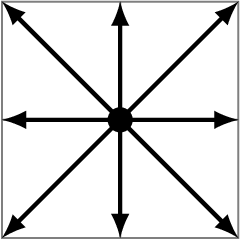}
        \begin{picture}(0,0)
        \put(-10, 62.5){\footnotesize (1, 1)}
        \put(-70, -10){\footnotesize (-1, -1)}
        \put(-55, -30){(a) D2Q9}
        \end{picture}
    }
    \hspace{2em}
    {
        \includegraphics[scale=1]{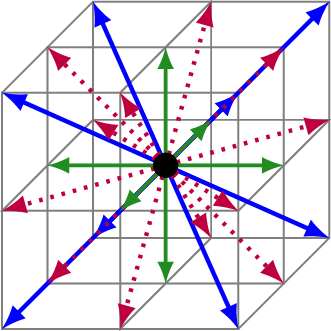}
        \begin{picture}(0,0)
        \put(-45, -10){\footnotesize (1, -1, 1)}
        \put(-100,-10){\footnotesize (-1, -1, 1)}
        \put(   0, 20){\footnotesize (1, -1, -1)}
        \put(   0, 75){\footnotesize (1, 1, -1)}
        \put(-105, 75){\footnotesize (-1, 1, -1)}
        \put(-75, -30){(b) D3Q27}
        \end{picture}
    }
    \vskip3em
    \caption{Velocity space representations in two- and three-dimensions using the standard D2Q9 and D3Q27 lattices respectively.
    }
    \label{fig:lattice-representation}
    \vskip-1.25cm
\end{figure}

Specifically, the discrete equilibrium is required to recover the following Maxwell-Boltzmann (MB) moments:
\begin{subequations}
	\label{eq:eqConstraints}
	\vspace{-1.5em}
	\begin{equation}
		\Big\langle f_i^{(eq)}, \big\lbrace 1, c_{i_\alpha}, c_i^2/2 \big\rbrace\Big\rangle = \left\lbrace \rho, \rho u_\alpha, E \right\rbrace\label{eq:conserved-quantitites}
	\end{equation}
	\vspace{-1cm}
    \begin{align}
        \Pi_{\alpha\beta}^{(eq)} =  \Pi_{\alpha\beta}^\text{MB} &= p\delta_{\alpha\beta} + \rho u_\alpha u_\beta\label{eq:eq-pressure-tensor}\\
        \vspace{-0.8em}q_\alpha^{(eq)} = q_\alpha^\text{MB} &= u_\alpha(E + p)\label{eq:eq-heatflux}
    \end{align}
	\vspace{-1em}
\end{subequations}

\noindent where $\left\langle \varphi_i, \psi_i\right\rangle = \sum\limits_i \varphi_i \psi_i$, the quantities $\rho$, $\rho u_\alpha$ and $E = \Pi_{\alpha\alpha}^{(eq)}/2 = (Dp + \rho u^2)/{2}$ are the mass, momentum, and energy density respectively, with $p=\rho\theta$ being the thermodynamic pressure given through the ideal gas law, where $\theta$ = $RT$ is the reduced temperature, and  $\Pi_{\alpha\beta}^{(eq)} = \left\langle f_i^{(eq)}, c_{i_\alpha}c_{i_\beta} \right\rangle$ and $q_\alpha^{(eq)} = \left\langle f_i^{(eq)}, \frac{1}{2}c_i^2 c_{i_\alpha} \right\rangle$ are the equilibrium contributions of the pressure tensor and heat flux vector.
Note that Eq.~(\ref{eq:eqConstraints}) corresponds to 8 and 13 linearly independent constraints on $f_i^{(eq)}$ in 2- and 3-dimensions.
Thus, one can completely describe the equilibrium state even on standard lattices (\Cref{fig:lattice-representation}) lying within the first Brillouin zone.
In contrast, on standard lattices, the description of the non-equilibrium populations, $f_i^{(neq)}$, and consequently the resulting out-of-equilibrium macrodynamics, requires more care.

We recall from kinetic theory that $f_i^{(neq)}$ must satisfy the so-called compatibility conditions (Eq.~\ref{eq:compatibility-condition}) and, should recover the NSF constitutive relations for the viscous stress tensor (Eq.~(\ref{eq:neq-pressure-tensor})) and heat flux vector (Eq.~\ref{eq:neq-heatflux}));
\begin{subequations}
	\begin{equation}
	    \Big\langle f_i^{(neq)}, \big\lbrace 1, c_{i_\alpha}, c_i^2/2 \big\rbrace\Big\rangle = 0, \label{eq:compatibility-condition}
	\end{equation}
    \label{eq:neqConstraints}
    \begin{align}
        \vspace{-1em}
        \hspace{-0.7425em}
        \Pi_{\alpha\beta}^{(neq)} = \Pi_{\alpha\beta}^\text{NSF} &= -\mu \bigg(\partial_\alpha u_\beta + \partial_\beta u_\alpha - \frac{2}{D}\partial_\chi u_\chi\delta_{\alpha\beta}\bigg)\label{eq:neq-pressure-tensor}\\
        \vspace{-1em}
        &q_\alpha^{(neq)} = q_\alpha^\text{NSF} = -\kappa \partial_\alpha \theta.\label{eq:neq-heatflux}
    \end{align}
\end{subequations}
\noindent The quantities $\mu$ and $\kappa$ are the dynamic viscosity and thermal conductivity respectively.

In order to evaluate Eq.~(\ref{eq:neqConstraints}), closed form expressions of $f_i^{(neq)}$ are needed; while newer regularization schemes prescribe analytical expressions \citep{jonnalagadda2023NHTB}, historically, $f_i^{(neq)}$ is simply computed as $f_i^{{(neq)}} = \Big(f_i - f_i^{(eq)}\Big)$.
However, since $f_i$ is a computationally evaluated quantity, such a representation is not amenable for a theoretical evaluation of Eq.~(\ref{eq:neqConstraints}).
Consequently, the following approach is adopted for enabling theoretical analyses:
Firstly, the compatibility conditions given by Eq.~(\ref{eq:compatibility-condition}) are \textit{assumed} to hold.
However, kinetic theory requires these conditions to be imposed on non-equilibrium descriptions of the distribution function \citep{amitagrawalbook} and, as evidenced by theoretical works on higher-order continuum transport equations \citep{o13, oBurnett, upendraAIP2023, upendraJFM2023}, such an assumption may not be valid even in the continuous phase space.
Secondly, Eqs.~(\ref{eq:neq-pressure-tensor}) and (\ref{eq:neq-heatflux}) are evaluated by coupling $\Pi_{\alpha\beta}^{(neq)}= \left\langle f_i^{(neq)}, c_{i_\alpha}c_{i_\beta} \right\rangle $ and $q_\alpha^{(neq)}= \left\langle f_i^{(neq)}, \frac{1}{2}c_i^2 c_{i_\alpha} \right\rangle $ to $f_i^{(eq)}$ via higher-order equilibrium moments $Q_{\alpha\beta\gamma}^{(eq)} = \left\langle f_i^{(eq)}, c_{i_\alpha}c_{i_\beta}c_{i_\gamma} \right\rangle$ and $R_{\alpha\beta\gamma\mu}^{(eq)} = \left\langle f_i^{(eq)}, c_{i_\alpha}c_{i_\beta}c_{i_\gamma}c_{i_\mu} \right\rangle$.
However, this coupling to higher order equilibrium moments has severe repercussions in that spurious numerical errors are introduced on standard lattices due to insufficient lattice isotropy.
In order to retain the computational advantage provided by LB on standard lattices, significant efforts have been made over the past three decades to eliminate these spurious errors \citep{ChenOhashiAkiyamaPRE1994, WagnerYeomansPRE1999, dellarPRE2002, dellarJCP2003, HaziKavranJPA2006, HaziJimenezCompFlu2006, KeatingValhalaYepezSoeValhalaPRE2007, dellarJCP2014, PrasianakisKarlin2007, PrasianakisKarlin2008, PrasianakisKarlinMantzarasBoulouchosPRE2009, SaadatBoschKarlinPRE2019, SaadatDorschnerKarlin2021}.
The current state-of-the-art incorporates non-local correction terms into the lattice update to exactly recover the desired thermohydrodynamics \citep{dellarJCP2014, PrasianakisKarlin2007, SaadatBoschKarlinPRE2019}.
However, it is noteworthy that this strategy is not only hard to generalize to different collision kernels and equilibrium representations, but is also detrimental to the parallel efficiency offered by LB on standard lattices.
Furthermore, this performance sacrifice is exacerbated in numerical implementations that incorporate advanced meshing paradigms such as, e.g., grid refinement, needed for addressing complex problems involving fluid flow.

In this work, we demonstrate that the recently proposed Onsager-Regularized (OReg) LB method \citep{ jonnalagaddaJHT2021, jonnalagaddaPRE2021, jonnalagadda2023NHTB} facilitates simulations of nonlinear NSF hydro-thermal transport phenomenon by alleviating the adverse standard lattice anisotropy effects without resorting to non-local corrections.
Specifically, we first present a generalized, assumption-free, and stencil-independent theoretical analysis of the OReg scheme via the Chapman-Enskog (CE) procedure \citep{succi2001lattice,kruger} that is valid for any equilibrium representation.
We then employ the so-called $\mathcal{O}(u^4)$ guided equilibrium  \citep{PrasianakisKarlin2007, PrasianakisKarlin2008, PrasianakisKarlinMantzarasBoulouchosPRE2009} and the $\mathcal{O}(u^2)$ polynomial equilibrium representation to obtain OReg kinetic models for the NSF equations on standard lattices.
Thereafter, we present numerical results for canonical quasi-one-dimensional problems demonstrating the improved accuracy obtained with the OReg model with the guided equilibrium, and, finally, present our conclusions.

\section{Onsager-Regularized LBM}

The OReg scheme uses the principles of linear irreversible thermodynamics and prescribes the non-equilibrium contribution of the distribution function at the NSF level in terms of viscous and thermal irreversible processes \citep{mahendra-thesis, mahendra2013onsager} that comply with Onsager's Symmetry Principle.
Here, for the sake of brevity, we restrict the discussion to isothermal flows of monatomic gases and direct the interested reader to Ref. \citep{jonnalagaddaPRE2021} and the references therein for a more detailed exposition.
In such a setting, the non-equilibrium thermodynamics representation of the NSF contribution of the distribution function is given as:
\vspace{-0.75em}
\begin{multline}
\label{eq:onsager-first-order}
f^{(1)}
=
f^{\text{OReg}}
=
- \tau\, f^\text{MB}
    \left(
        C_\alpha C_\beta - \frac{C^2}{D}  \delta_{\alpha\beta}
    \right)
    \\
    \left(
       \frac{\partial_\beta u_\alpha + \partial_\alpha u_\beta}{2\theta}
    \right).
\vspace{-0.5em}
\end{multline}

\noindent
Here, $f^\text{MB}$ is the continuous Maxwell-Boltzmann equilibrium distribution function and $C_\alpha = \left(c_\alpha - u_\alpha\right)$ is the peculiar velocity.
Note that Eq.~(\ref{eq:onsager-first-order}) can be recast into the same form \citep{jonnalagaddaPRE2021} that is obtained from a CE expansion of the continuous, force-free, integro-differential Boltzmann equation and thus recovers the isothermal Navier-Stokes (NS) equations exactly in the non-discrete setting.
Here we remark that, much like the entropic LB which connects the discrete equilibrium and path length to kinetic theory via the discrete H-Theorem \citep{hosseini2023ELBMreview}, the OReg formulation explicitly connects the description of the non-equilibrium lattice populations to kinetic theory via non-equilibrium thermodynamic descriptions.
This is done by simply projecting Eq.~(\ref{eq:onsager-first-order}) onto lattice stencils by representing $f^\text{MB}$ and $C_\alpha$ as $f_i^{(eq)}$ and $C_{i_\alpha} = \left(c_{i_\alpha} - u_\alpha\right)$ respectively to obtain:
\vspace{-0.5em}
\begin{multline}
\label{eq:onsager-lattice-discretized}
f_i^{\text{OReg}}
=
- \tau\, f^{(eq)}
    \left(
        C_{i_\alpha} C_{i_\beta} - \frac{C^2_i}{D}  \delta_{\alpha\beta}
    \right)
    \\
    \left(
       \frac{\partial_\beta u_\alpha + \partial_\alpha u_\beta}{2\theta}
    \right).
\end{multline}
Thereafter, the derivative terms appearing in Eq.~(\ref{eq:onsager-lattice-discretized}) are locally evaluated by using the definition of the second-order trace-free stress tensor \citep{jonnalagaddaJHT2021, jonnalagaddaPRE2021, jonnalagadda2023NHTB} to generically represent the OReg populations as:
\vspace{-1em}
\begin{multline}
    \label{eq:Oregf1}
    f_i^{\text{OReg}}
    =
    \frac{f_i^{(eq)}}{2\rho \theta^2}
        \left(
            C_{i_\alpha} C_{i_\beta} - \frac{C_i^2}{D}  \delta_{\alpha\beta}
        \right)
        \\[-0.5em]
        \sum\limits_{k=1}^{N}
        \Big(
            c_{k_\alpha}c_{k_\beta} - \frac{c_k^2}{D} \delta_{\alpha\beta} 
        \Big)f_k^{(neq)}
        .
\end{multline}

\vspace{-0.75em}
\noindent
Note that Eq.~(\ref{eq:Oregf1}) is agnostic to the lattice stencil as well as the spatial discretization and therefore yields a fully local formulation.
It is also noteworthy that, as opposed to earlier works \citep{jonnalagaddaJHT2021, jonnalagaddaPRE2021, jonnalagadda2023NHTB}, the OReg scheme as presented in Eq.~(\ref{eq:Oregf1}) is generalized to include temperature dependence and is a significantly simpler representation of the scheme valid for monatomic gases.
We highlight that the presence of $f_i^{(neq)}$ in Eq.~(\ref{eq:Oregf1}) allows the OReg scheme to be interpreted as a one-step predictor-corrector method where the prediction $f_i^{(1)} \approx f_i^{(neq)}$ is corrected to $f_i^{(1)}= f_i^{\text{OReg}}$ evaluated using Eq.~(\ref{eq:Oregf1}).
The predictor $f_i^{(neq)}$ can be evaluated through any existing regularization scheme \citep{jonnalagadda2023NHTB}; however, for the sake of simplicity, we use $f_i^{(neq)}=\left(f_i - f_i^{(eq)}\right)$.
Finally, in line with regularized LBM schemes, the OReg lattice update is realized by reworking Eq.~(\ref{eq:lattice-update-std}) as:
\vspace{-1em}
\begin{equation}
	\label{eq:lattice-update}
	f_i(x_\alpha + c_{i_\alpha}\Delta t, t + \Delta t) = f_i^{(eq)}(x_\alpha, t) + \Big(1-\frac{1}{\tau}\Big)f_i^{\text{OReg}}(x_\alpha, t).
\end{equation}

Before proceeding, we remark that Eq.~(\ref{eq:Oregf1}) is strikingly similar to those obtained with the projected regularized scheme \citep{LATT2006} and, is visually almost identical to the Galilean Invariant Filtered (GIFR) collision model \citep{chen2020filtered}; this is an expected outcome since these schemes aim to recover the same macroscopic equations.
However, it is important to note that the projected regularized scheme, along with systematic improvements such as the recursively regularized \citep{malaspinasArXiV,coreixasRecReg}, hybrid-recursively regularized \citep{fengHybridRecreg2019} and GIFR schemes, are \textit{functional approximations} of the lattice-discretized non-equilibrium distribution function expressed in a Hermitian basis \citep{jonnalagadda2023NHTB}.
In contrast, the OReg scheme, represented by Eq.~(\ref{eq:Oregf1}), is simply a fully local lattice-discretized representation of the non-equilibrium distribution function \citep{jonnalagadda2023NHTB}.
It is, therefore, reasonable to expect that the OReg scheme would yield improved stability characteristics as compared to other regularization strategies; indeed, previous work \citep{jonnalagaddaPRE2021} has demonstrated such a behaviour for flows with large Reynolds (Re) and Mach (Ma) numbers on standard lattices.

\vspace{1em}
\section{Hydrodynamic limit using the Chapman-Enskog multiscale expansion}

\vspace{-0.5em}
\subsection{General outcomes}
\vspace{-0.5em}

We now briefly describe the procedure to obtain the hydrodynamic limit of single relaxation time LB schemes through the Chapman-Enskog (CE) multiscale expansion \citep{succi2001lattice,kruger}.
The quantities appearing in Eq.~(\ref{eq:lattice-update-std}) are expressed in terms of a perturbation series as $f_i = f_i^{(0)} + \epsilon f_i^{(1)} + \epsilon^2 f_i^{(2)} +\mathcal{O}(\epsilon^3)$, $\partial_t = \partial_t^{(0)} + \epsilon\partial_t^{(1)} + \mathcal{O}(\epsilon^2)$ and $\partial_\alpha = \epsilon\partial_\alpha^{(1)}$, where the perturbation parameter, $\epsilon$, corresponds to the Knudsen number (Kn). 
Subsequently, after a series of algebraic manipulations, during which it is revealed that $f_i^{(0)} = f_i^{(eq)}$, the following $\mathcal{O}(\epsilon^3)$ equations are recovered:
\begin{widetext}
    \vspace{-2em}
    \begin{subequations}
	\label{eq:CEmacroscopicEqs}
        \begin{align}
            \partial_t \varphi_i^{(eq)} + \partial_\alpha \varphi_{i_\alpha}^{(eq)}
            =
            -\frac{\epsilon}{\tau}\varphi_i^{(1)}
            -\frac{\epsilon^2}{\tau}\varphi_i^{(2)}
            -\epsilon^2
            \Bigg\lbrace
                \partial_t^{(1)}
                \Bigg[
                    \bigg( 1 - \frac{\Delta t}{2\tau} \bigg)
                    \varphi_i^{(1)}
                \Bigg]
                +
                \partial_\alpha^{(1)}
                &
                \Bigg[
                    \bigg( 1 - \frac{\Delta t}{2\tau} \bigg)
                    \varphi_{i_\alpha}^{(1)}
                \Bigg]
            \Bigg\rbrace,
            \label{eq:CEcontinuity}
            \\
            \partial_\alpha \varphi_{i_\alpha}^{(eq)}
            +
            \partial_\alpha
            \underbrace{
                \left\langle
                    f_i^{(eq)}, c_{i_\alpha}c_{i_\beta}
                \right\rangle}_{\Pi_{\alpha\beta}^{(eq)}}
            +
            \,
            \partial_\alpha
            \Bigg[
                \epsilon
                \bigg( 1 - \frac{\Delta t}{2\tau} \bigg)
                \underbrace{
                    \left\langle
                        f_i^{(1)}, c_{i_\alpha}c_{i_\beta}
                    \right\rangle}_{\Pi_{\alpha\beta}^{(1)}}
            \Bigg]
            =
            -\frac{\epsilon}{\tau}\varphi_{i_\alpha}^{(1)}
            -\frac{\epsilon^2}{\tau}\varphi_{i_\alpha}^{(2)}
            &
            -\epsilon^2
            \partial_t^{(1)}
            \Bigg[
                \bigg( 1 - \frac{\Delta t}{2\tau} \bigg)
                \varphi_{i_\alpha}^{(1)}
            \Bigg]
            \label{eq:CEmomentum},
        \end{align}
    \end{subequations}
    \vspace{-1em}
\end{widetext}
\noindent
where $\Big\lbrace\varphi_i^{(eq)}, \varphi_i^{(1)}, \varphi_i^{(2)} \Big\rbrace = \left\langle \Big\lbrace f_i^{(eq)}, f_i^{(1)},f_i^{(2)} \Big\rbrace , 1 \right\rangle$ and $\Big\lbrace\varphi_{i_\alpha}^{(eq)},\varphi_{i_\alpha}^{(1)},\varphi_{i_\alpha}^{(2)}\Big\rbrace = \left\langle \Big\lbrace f_i^{(eq)}, f_i^{(1)}, f_i^{(2)}\Big\rbrace, c_{i_{\alpha}}\right\rangle$.
Note that, with $f_i^{(0)} = f_i^{(eq)}$, the resultant non-equilibrium populations become $f_i^{(neq)} = \epsilon f_i^{(1)} + \epsilon^2 f_i^{(2)} + \mathcal{O}(\epsilon^3)$, and upon introducing the constraints of Eq.~(\ref{eq:eqConstraints}) and Eq.~(\ref{eq:neqConstraints}), Eq.~(\ref{eq:CEcontinuity}) directly yields the equation of mass conservation, while Eq.~(\ref{eq:CEmomentum}), pending closure of $\Pi_{\alpha\beta}^{(1)}$, approximates the NS momentum conservation equations; as alluded to earlier, this closure is typically performed via $Q^{(eq)}_{\alpha\beta\gamma}$.

\vspace{-1.75em}
\subsection{Onsager-Regularized macrodynamics for a generic equilibrium distribution function}
\vspace{-1em}

\noindent
Now that the generalized outcome of the CE expansion at the NS level is available in Eq.~(\ref{eq:CEmacroscopicEqs}), we move on to obtain the macroscopic limit of the OReg scheme for a generic equilibrium representation.
Note that such a generic representation need not satisfy Eq.~(\ref{eq:eqConstraints}) in its entirety for recovering athermal/isothermal hydrodynamics.
Indeed, on standard lattices, only the so-called consistent equilibrium \citep{consistentLBM}, when operated at the standard lattice reference temperature $\theta_0 = 1/3$, satisfies Eq.~(\ref{eq:eqConstraints}) completely.
Therefore, we introduce $\Phi_i' = \Phi_i^{(eq)} - \Phi_i^\text{MB}$ as the  generic deviation of an equilibrium moment $\Phi_i^{(eq)}$ from its MB counterpart $\Phi_i^\text{MB}$ and $\Pi_{\alpha\beta}'$, $q_\alpha'$, $Q_{\alpha\beta\gamma}'$ and $R_{\alpha\beta\gamma\mu}'$ as the respective deviations of the equilibrium pressure tensor, heat flux vector and the third- and fourth-order moments respectively.
Note that the conserved quantities are expected to not have any deviations, i.e., $\rho' = (\rho u_\alpha)' = 0$.
Thereafter, since we are interested in NS level thermohydrodynamics, we assume that the compatibility conditions hold for the Burnett and higher order contributions; thus $\varphi_i^{(2)}$ and $\varphi_{i_\alpha}^{(2)}$ are always taken to be zero while $\varphi_i^{(1)}$, $\varphi_{i_\alpha}^{(1)}$ are evaluated explicitly using with $f_i^{(1)} = f_i^\text{OReg}$ given in Eq.~(\ref{eq:onsager-lattice-discretized}).
Note that this is the first deviation from the standard procedure followed in the LB literature which assumes that Eq.~(\ref{eq:compatibility-condition}) holds for all levels of approximation in $\epsilon$.
Explicitly, we obtain:
\begin{widetext}
	\vspace{-0.75cm}
    \begin{subequations}
    	    \label{eq:Oreg-compatibility-conditions}
        \begin{align}
            \varphi_i^{(1)|\text{OReg}}
            =
            \Big\langle f_i^\text{OReg}, 1\Big\rangle
            =
            -\frac{\tau}{2\theta}
            \left(
                \partial_\beta^{(1)} u_\alpha + \partial_\alpha^{(1)} u_\beta
            \right)
            \left(
                \Pi_{\alpha\beta}' - \frac{\delta_{\alpha\beta}}{2D}\Pi_{\chi\chi}'
            \right)
            \phantom{~~~~~~~~~~}
            \label{eq:oreg-mass-compatibility}
            \\[-0.15em]
            \varphi_{i_{\gamma}}^{(1)|\text{OReg}}
            {
            \,
            =
            \Big\langle f_i^\text{OReg}, c_{i_\gamma}\Big\rangle}
            =
            -\frac{\tau}{2\theta}
            \left(
                \partial_\beta^{(1)} u_\alpha + \partial_\alpha^{(1)} u_\beta
            \right)
            \left[
                Q_{\alpha\beta\gamma}' - u_\alpha\Pi_{\beta\gamma}' - u_\beta\Pi_{\gamma\alpha}' - \frac{\delta_{\alpha\beta}}{D} \left(q_\gamma' - \frac{u_\gamma}{2}\Pi_{\chi\chi}'\right)
            \right],
            \label{eq:oreg-mom-compatibility}
        \end{align}
    \end{subequations}
    \vspace{-0.5cm}
\end{widetext}
\noindent
where the derivative terms have the superscript (1) since $f_i^\text{OReg}$ is of $\mathcal{O}(\epsilon)$.
The next deviation from the standard procedure is in closing Eq.~(\ref{eq:CEmomentum}) which is achieved through directly evaluating $\Pi_{\alpha\beta}^{(1)}$ using $f_i^\text{OReg}$.
This direct evaluation yields:
\begin{widetext}
    \vspace{-0.75cm}
    \begin{align}
        \Pi_{\alpha\beta}^{(1)|\text{OReg}}
        =
        \left\langle f_i^\text{OReg}, c_{i_\alpha}c_{i_\beta}\right\rangle
        =
         \underbrace{-\tau\rho\theta \bigg(\partial^{(1)}_\alpha u_\beta + \partial^{(1)}_\beta u_\alpha - \frac{2}{D}\partial^{(1)}_\chi u_\chi\delta_{\alpha\beta}\bigg)}_{\Pi_{\alpha\beta}^\text{NSF}}
         \phantom{~~~~~~~~~~~~~~~~~~~}
         \nonumber
         \\
         \underbrace{
            -\frac{\tau}{2\theta}\left(\partial_\mu^{(1)} u_\gamma + \partial_\gamma^{(1)} u_\mu\right)
            \Big[
                R_{\alpha\beta\gamma\mu}'
                -
                u_\gamma Q_{\alpha\beta\mu}'
                -
                u_\mu Q_{\alpha\beta\gamma}'
                +
                u_\gamma u_\mu \Pi_{\alpha \beta}'
                -
                \frac{\delta_{\gamma\mu}}{D}
                \big(
                    R_{\alpha\beta\chi\chi}'
                    -
                    2 u_\chi Q_{\alpha\beta\chi}'
                    +
                    u_\chi^2\Pi_{\alpha\beta}'
                \big)
            \Big]
        }_{\left(\Pi_{\alpha\beta}^{(neq)}\right)'}
        ,
        \label{eq:oreg-pressure-tensor}
    \end{align}
	\vspace{-0.5cm}
\end{widetext}
\noindent where $\left(\Pi_{\alpha\beta}^{(neq)}\right)'$ is the error contribution to the NSF stress tensor.
Here, we highlight that, while Eqs.~(\ref{eq:Oreg-compatibility-conditions}) and (\ref{eq:oreg-pressure-tensor}) can be simplified depending on the constraints on $f_i^{(eq)}$, spurious contributions from $Q_{\alpha\beta\gamma}'$ and $R_{\alpha\beta\gamma\mu}'$ are still retained on standard lattices.
Nevertheless, it is important to note that Eq.~(\ref{eq:oreg-pressure-tensor}) recovers the NSF stress tensor, albeit with the erroneous $\Big(\Pi_{\alpha\beta}^{(neq)}\Big)'$ contribution, irrespective of the employed equilibrium $f_i^{(eq)}$.
It is worth noting that the derivative terms pre-multiplying the error terms increase the magnitude of $\Big(\Pi_{\alpha\beta}^{(neq)}\Big)'$ by $\mathcal{O}(u)$.
Thus, the overall equilibrium-dependent spurious contributions brought in by $Q_{\alpha\beta\gamma}'$ and $R_{\alpha\beta\gamma\mu}'$ are diluted by at least one order of magnitude and, consequently, may be small enough to be ignored.
In contrast, when the closure of Eq.~(\ref{eq:CEmomentum}) is performed via $Q_{\alpha\beta\gamma}^{(eq)}$, only the consistent equilibrium of Ref. \citep{consistentLBM} yields a trace-free NSF stress tensor.

\subsection{Onsager-Regularized macrodynamics using the Guided Equilibrium}

\noindent
As an explicit illustration of the above arguments, we first consider the OReg scheme coupled with the guided equilibrium distribution representation \citep{PrasianakisKarlin2007, PrasianakisKarlin2008, PrasianakisKarlinMantzarasBoulouchosPRE2009}:
\begin{equation}
    \label{eq:guidedEq}
    f_i^{(eq)|\text{G}}= \rho\prod\limits_{\alpha=x,y} \frac{(1-2c^2_{i_\alpha})}{2^{c^2_{i_\alpha}}}\left[c^2_{i_\alpha} - 1 + c_{i_\alpha}u_\alpha + u^2_\alpha + \theta\right].
\end{equation}

\noindent
Note that Eq.~(\ref{eq:guidedEq}), defined for the D2Q9 lattice obeys six out of the eight constraints imposed on $f_i^{(eq)}$ with Eq.~(\ref{eq:eq-heatflux}) not being obeyed.
Using Eq.~(\ref{eq:guidedEq}) in Eq.~(\ref{eq:onsager-lattice-discretized}), we  explicitly obtain the OReg non-equilibrium populations which we denote using $f_i^{\text{OReg}|\text{G}}$.
Thereafter, substituting $f_i^{\text{OReg}|\text{G}}$ into Eq.~(\ref{eq:oreg-pressure-tensor}) gives us the OReg NS pressure tensor using the guided equilibrium as:
\begin{widetext}
\vspace{-1em}
\begin{subequations}
	\begin{equation}
	    \Pi_{\alpha\beta}^{(1)|\text{OReg}|\text{G}}
        =
        -\mu^{\text{OReg}|\text{G}}_{\alpha\beta}
        \bigg(\partial_\alpha u_\beta + \partial_\beta u_\alpha - \partial_\chi u_\chi\delta_{\alpha\beta}\bigg)
        +
        \left(\Pi_{\alpha\beta}^{(neq)|\text{G}}\right)'
        \label{eq:guidedEq-oreg-neq-pressure}	
	\end{equation}

    \begin{align}
        \text{where, ~~}\mu^{\text{OReg}|\text{G}}_{\alpha\beta}
        =
        &
        \tau\rho\theta
        \begin{cases}
            \frac{1}{2\theta^2}(\theta - u_\alpha^2)(1-\theta-u_\alpha^2) &\mbox{ if } \alpha=\beta \\
            \phantom{\frac{1}{2\theta^2}(\theta - u_\alpha^2)} 1 &\mbox{ otherwise,}
        \end{cases}
        \label{eq:guidedEq-oreg-viscosity}
        \\
        \text{and, ~~}
        \left(\Pi_{\alpha\beta}^{(neq)|\text{G}}\right)'
        &
        =
        \begin{cases}
            \phantom{\rho\frac{\tau}{2\theta}u_x u_y (u_x^2 -} 0 &\mbox{if } \alpha=\beta \\
            \rho\frac{\tau}{2\theta}u_x u_y (u_x^2 - u_y^2) (\partial_x u_x - \partial_y u_y) \sim \mathcal{O}(u^6)&\mbox{otherwise.}
        \end{cases}
        \label{eq:guidedEq-oreg-pressure-error}
    \end{align}
\end{subequations}
\vspace{-0.5em}
\end{widetext}

\vspace{-2em}
\noindent
It is remarkable to note from Eq.~(\ref{eq:guidedEq-oreg-pressure-error}) that the diagonal components of the non-equilibrium pressure tensor are naturally error-free while the off-diagonal components have an $\mathcal{O}(u^6)$ error (recall that $\tau \propto u$).
Thus, the contribution of $\left(\Pi_{\alpha\beta}^{(neq)|\text{G}}\right)'$ to $\Pi_{\alpha\beta}^{\text{OReg}|\text{G}}$ is indeed negligible and can be safely ignored.
Now, with a vanishing $\left(\Pi_{\alpha\beta}^{(neq)|\text{G}}\right)'$, Eq.~(\ref{eq:guidedEq-oreg-neq-pressure}) explicitly yields a trace-less second-order pressure tensor having a velocity and temperature dependant dynamic viscosity of $\mu^{\text{OReg}|\text{G}}_{\alpha\beta}$ as shown in Eq.~(\ref{eq:guidedEq-oreg-viscosity}).
It is through $\mu^{\text{OReg}|\text{G}}_{\alpha\beta}$ that the OReg scheme compensates for the insufficient lattice isotropy of the D2Q9 lattice by intrinsically modifying the lattice viscosity of the diagonal components of $\Pi_{\alpha\beta}^{\text{OReg}|\text{G}}$.

Next, we evaluate Eqs.~(\ref{eq:oreg-mass-compatibility}) and (\ref{eq:oreg-mom-compatibility}) using the guided equilibrium to obtain:
\begin{widetext}
\vspace{-0.5cm}
\begin{subequations}
    \begin{align}
	    \varphi_i^{(1)|\text{OReg}|\text{G}}
        =\,
        \Big\langle f_i^{\text{OReg}|\text{G}}, 1\Big\rangle
        =
        0,
        \phantom{~~~~~~~~~~~~~~~~~~~~~~~~~~~~~~~~~~}
        \label{eq:guidedEq-oreg-mass-compatilbility}
        \\
        \varphi_{i_\alpha}^{(1)|\text{OReg}|\text{G}}
        =\,
        \Big\langle f_i^{\text{OReg}|\text{G}}, c_{i_{\alpha}}\Big\rangle
        =
        \frac{\tau\rho u_\alpha}{\theta}
        \Big(
            u_\alpha^2 + 3\theta - 1
        \Big)
        \bigg(
            \partial_\alpha u_\alpha - \frac{1}{2}\partial_\chi u_\chi
        \bigg)
        =
        \begin{cases}
            \mathcal{O}(u^5) &\text{if } \theta = 1/3\\
            \mathcal{O}(u^3) &\text{otherwise}
        \end{cases}.
        \label{eq:guidedEq-oreg-mom-compatilbility}
    \end{align}
\vspace{-1em}
\end{subequations}
\end{widetext}
\noindent
From Eq.~(\ref{eq:guidedEq-oreg-mass-compatilbility}), it can be seen that Eq.~(\ref{eq:oreg-mass-compatibility}) vanishes with the OReg scheme and the guided equilibrium as required by the compatibility conditions; in contrast, $\varphi_{i_\alpha}^{(1)|\text{OReg}|\text{G}}$ is non-zero with errors having magnitudes as shown in  Eq.~(\ref{eq:guidedEq-oreg-mom-compatilbility}).
Consequently, due to the $(1/2\tau)$ factor in Eq~(\ref{eq:CEmacroscopicEqs}), the corresponding mass and momentum conservation equations contain $\mathcal{O}(u^4)$/$\mathcal{O}(u^2)$ errors depending on the value of the isothermal temperature.
Thus, with the OReg scheme and the guided equilibrium, one can conduct correction-free isothermal NS simulations with $\mathcal{O}(u^3)$/$\mathcal{O}(u)$ accuracy at the lattice reference temperature and arbitrary lattice temperatures, respectively.

In order to isolate the accuracy gained by using OReg scheme from those brought about by using the guided equilibrium, we consider the momentum equation obtained from the CE expansion of the Lattice-BGK scheme with the guided equilibrium from Ref. \citep{PrasianakisKarlin2007}:

\vspace{-1em}
\begin{widetext}
\vspace{-2em}
	\begin{subequations}
		\label{eq:guided-lbgk-CE}
		\begin{align}
			\partial_t(\rho u_\alpha)
			\,+\,
			\partial_\beta\Pi_{\alpha\beta}^\text{MB}
			+
			\partial_\beta
			\bigg[
				-
				\tau\rho\theta 
				\bigg(1-\frac{\Delta t}{2\tau}\bigg)
				\bigg(\partial_\alpha u_\beta + \partial_\beta u_\alpha - \partial_\chi u_\chi\delta_{\alpha\beta} \bigg)
			\bigg]
			=
			-\partial_\gamma\bigg[\bigg(1-\frac{\Delta t}{2\tau}\bigg)\left(\Pi_{\alpha\gamma}^{(neq)}\right)'\bigg],
			\label{eq:guided-lbgk-momEq}			
			\\[-1em]
			\intertext{where,\vspace{-1em}}
			\partial_\gamma\left(\Pi_{\alpha\gamma}^{(neq)}\right)'
			=
			\mp
			\frac{\tau}{2}
			\partial_\alpha
			\Bigg\lbrace
				\partial_x
				\Big[
					\rho u_x\big(1-3\theta\big)
					-
					\rho^2 u_x^3
				\Big]
				-
				\partial_y
				\Big[
					\rho u_y\big(1-3\theta\big)
					-
					\rho^2 u_y^3
				\Big]
			\Bigg\rbrace
			\sim
			\begin{cases}
				\mathcal{O}(u^4) \hspace{0.25cm}\text{if } \theta = 1/3 \\
				\mathcal{O}(u^2) \hspace{0.25cm}\text{otherwise} \\
			\end{cases}			
			,
			\label{eq:guided-lbgk-momEq-error}
		\end{align}
	\vspace{-1em}
	\end{subequations}
\end{widetext}

\vspace{-1em}
\noindent
where Eq.~(\ref{eq:guided-lbgk-momEq-error}) takes the negative value if $\alpha = x$ and vice-versa if $\alpha = y$ respectively.
It can be seen that, with the $1/2\tau$ factor from Eq.~(\ref{eq:guided-lbgk-momEq}), Eq.~(\ref{eq:guided-lbgk-momEq-error}) culminates into an $\mathcal{O}(u^3)$ and $\mathcal{O}(u)$ representation of the momentum equation when $\theta = 1/3$ and $\theta \neq 1/3$ respectively for the Lattice-BGK scheme used with the guided equilibrium.
However, we highlight that, since Eq.~(\ref{eq:guided-lbgk-momEq}) is derived under the assumption that Eq.~(\ref{eq:compatibility-condition}) holds, there may be additional unaccounted errors that remain.
In contrast, under the same assumptions, the OReg scheme with the guided equilibrium yields a negligible error irrespective of the lattice temperature with $\left(\Pi_{\alpha\beta}^{(neq)|\text{G}}\right)' \sim \mathcal{O}(u^6)$.
Thus, if the compatibility conditions are assumed to hold, the OReg scheme used with the guided equilibrium exactly models the NSF momentum equations with $\mu_{\alpha\beta}^{\text{OReg}|\text{G}}$ compensating for the lattice anisotropy.
We highlight that it is \textit{only} due to the explicit evaluation of the compatibility condition $\Big\langle f_i^{\text{OReg}|\text{G}}, c_{i_\alpha}\Big\rangle$ that the OReg scheme yields $\mathcal{O}(u^4)$ and $\mathcal{O}(u^2)$ accurate momentum representations when $\theta = 1/3$ and $\theta \neq 1/3$ respectively.
Note that this is still an order of magnitude better than the BGK/guided equilibrium formulation.

\subsection{Onsager-Regularized macrodynamics using the second order polynomial equilibrium}

As a further illustration, we consider the OReg scheme when used with the popular $\mathcal{O}(u^2)$ polynomial form of the equilibrium distribution commonly used at the lattice reference temperature $\theta_0 = 1/3$:
\newpage
\begin{equation}
	f_i^{(eq)|\mathcal{O}(u^2)} = \rho\, W_i \left[1 + \frac{u_\alpha c_{\alpha}}{\theta_0} + \frac{(u_\alpha c_{\alpha})^2}{2\theta_0^2}\right]
\end{equation}
\noindent where the weights $W_i$ for the D2Q9 lattice are obtained as tensor products of $W_{\pm 1} = \theta_0/2$ and $W_0 = (1-\theta_0)$.
Evaluating Eqs. (\ref{eq:Oreg-compatibility-conditions}) and (\ref{eq:oreg-pressure-tensor}) with $f_i^{\text{OReg}|\mathcal{O}(u^2)}$ yields:

\begin{widetext}
\begin{subequations}
	\vspace{-1em}
	\label{eq:2o-poly-eq-oreg-results}
	\begin{align}
		\varphi^{(1)|\text{OReg}|\mathcal{O}(u^2)}_i
        =
        \left\langle
            f_i^{\text{OReg}|\mathcal{O}(u^2)}, 1
        \right\rangle
        =
        0,
        \label{eq:2o-poly-eq-mass-compatibility-condition}
        \\
        \varphi^{(1)|\text{OReg}|\mathcal{O}(u^2)}_{i_\alpha}
        =
        \left\langle
            f_i^{\text{OReg}|\mathcal{O}(u^2)}, c_{i_\alpha}
        \right\rangle
        =
        \tikzmark{V1}
        \frac{\tau\rho}{2\theta_0} u_\alpha
        \Big[
        		(u_x^2 - u_y^2)(\partial_x u_x - &\partial_y u_y)
        		+
        		2 u_x u_y(\partial_y u_x - \partial_x u_y)
        \Big],
        \tikzmark{V2}
        \label{eq:2o-poly-eq-mom-compatibility-condition}
    \end{align}
\end{subequations}
\vspace{-1em}
\begin{align}
	\text{and, ~~}\left(\Pi_{\alpha\beta}^{(neq)|\mathcal{O}(u^2)}\right)'
	=
	-
	\underbrace{
        u_\beta\,\varphi^{(1)|\text{OReg}|\mathcal{O}(u^2)}_{i_\alpha}
    }_{\mathcal{O}(u^6)}
    -
    \underbrace{
        \frac{\tau\rho}{2\theta_0}
        u_\alpha
        u_\beta
        (
        		\partial_x u_x
        		-
        		\partial_y u_y
        )
    }_{\mathcal{O}(u^4)}
    \times
    \begin{cases}
    		\pm 1 &\mbox{if } \alpha = \beta = x/y \\
    		\phantom{-}0 &\mbox{if } \alpha \neq \beta
    \end{cases}
	.
    \label{eq:2o-poly-eq-oreg-pressure-error}
\end{align}
\end{widetext}
\noindent
Eq.~(\ref{eq:2o-poly-eq-mom-compatibility-condition}) shows that the error in the first order compatibility condition obtained using $f_i^{\text{OReg}|\mathcal{O}(u^2)}$ has a magnitude of $\mathcal{O}(u^5)$ and can be safely ignored while from Eq.~(\ref{eq:2o-poly-eq-oreg-pressure-error}) the error $\left(\Pi_{\alpha\beta}^{(neq)|\mathcal{O}(u^2)}\right)' \sim \mathcal{O}(u^4)$.
Correspondingly, the momentum equations from the OReg scheme using $f_i^{\text{OReg}|\mathcal{O}(u^2)}$ are $\mathcal{O}(u^3)$ accurate.
Again, the OReg scheme yields an improvement in accuracy of an order of magnitude as compared to the bare BGK scheme.

\section{Numerical Benchmarks}

We now proceed to numerically benchmark the OReg scheme used with the guided equilibrium for two canonical quasi-one-dimensional problems.
First the classical linear benchmark case of a decaying shear wave \citep{PrasianakisKarlinMantzarasBoulouchosPRE2009, dellarJCP2014, SaadatDorschnerKarlin2021} is considered.
In the absence of any spurious numerical errors, observed statistics for waves described by a wave vector $\bm{k} = m \hat{e}_{\|} + n \hat{e}_{\perp}$, where $\hat{e}_{\|}$ and $\hat{e}_{\perp}$ are unit vectors parallel and perpendicular to the wave, theoretically display an exponential time decay.
Here we take the amplitude of the $x$-velocity component, $u^{max}_x(t)$, as the statistic of interest and compute the numerical kinematic viscosity, $\widetilde{\nu}$, through the following curve fitting expression: $u^{max}_x(t)\propto \exp\left(-|\bm{k}^2|\widetilde{\nu}t \right)$.
We consider two situations, namely an axis-aligned and a $\pi/4$-rotated wave, respectively.
Both cases are initialized with a unit density.
The initial velocity field for the axis-aligned case is given as:
\begin{figure}[h]
    \centering
    {
    \includegraphics[scale=1]{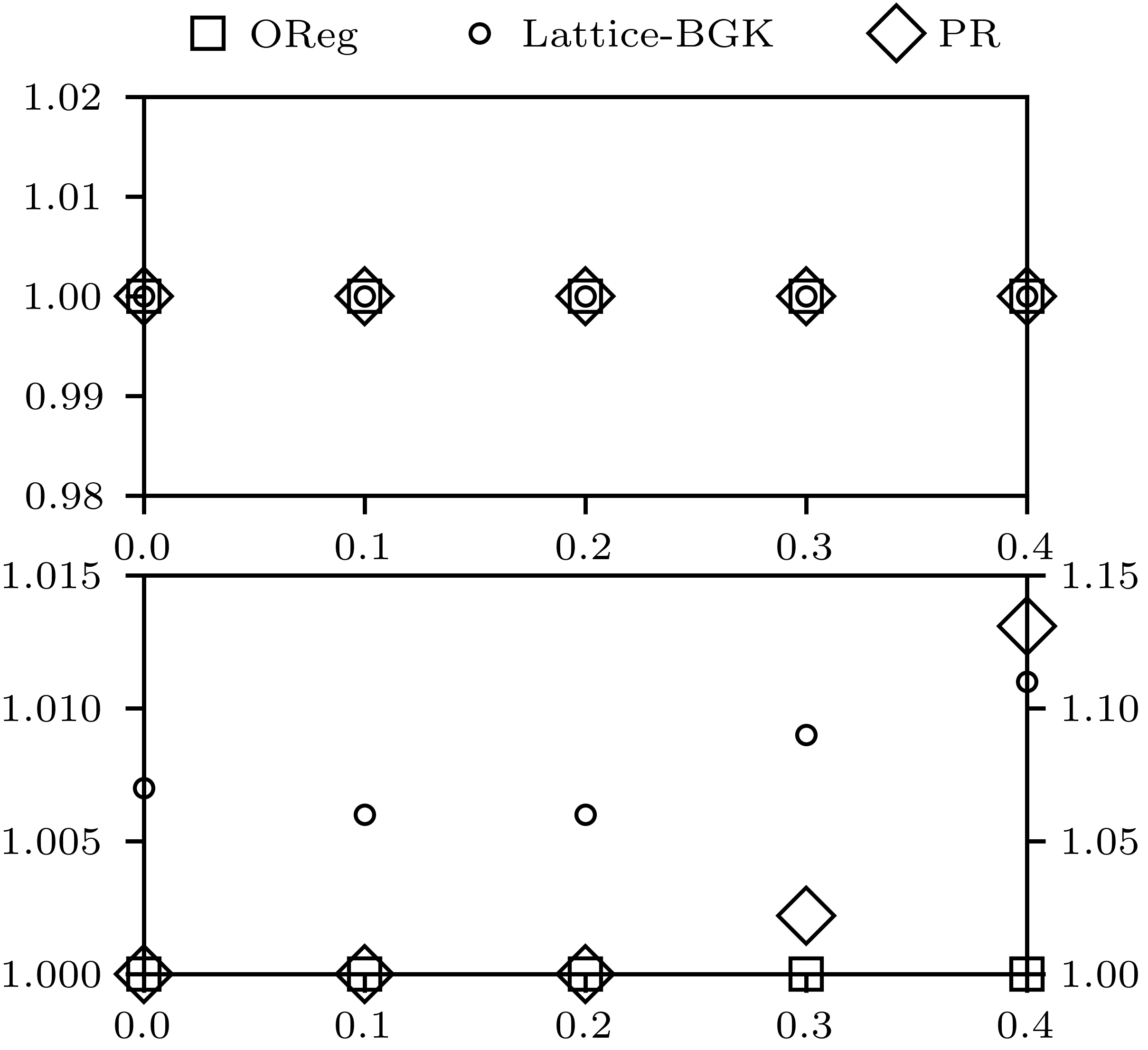}
    \begin{picture}(0,0)
        \put(-192.5,170){\footnotesize(a)}
        \put(-192.5,80){\footnotesize(b)}
        \put(-140,-6){\footnotesize Mach number}
        \put(-232.5,47.5){ \rotatebox{90}{$\widetilde{\nu}/\nu$}}
        \put(-7,40){ \rotatebox{90}{$\widetilde{\nu^\text{PR}}/\nu$}}
        \put(-232.5,135){ \rotatebox{90}{$\widetilde{\nu}/\nu$}}
    \end{picture}
    }
\vskip1em
\caption{\label{fig:wavedecay}Comparison of the numerically computed and physically imposed fluid viscosities, $\widetilde{\nu}$ and $\nu$, for a decaying shear wave at different Mach numbers obtained with the OReg, Lattice-BGK and Projected Regularized (PR) LB schemes using the guided equilibrium on the D2Q9 lattice. For the axis-aligned shear wave (panel (a)) all schemes correctly model the viscous dissipation as demonstrated by $\widetilde{\nu}/\nu$ = 1 while for the $\pi/4$ rotated wave (panel (b)) only the OReg scheme recovers the imposed viscous dissipation rate.}
 \vskip-1em
\end{figure}

\begin{figure*}[t]
\centering
    \centering
    \includegraphics[scale=1]{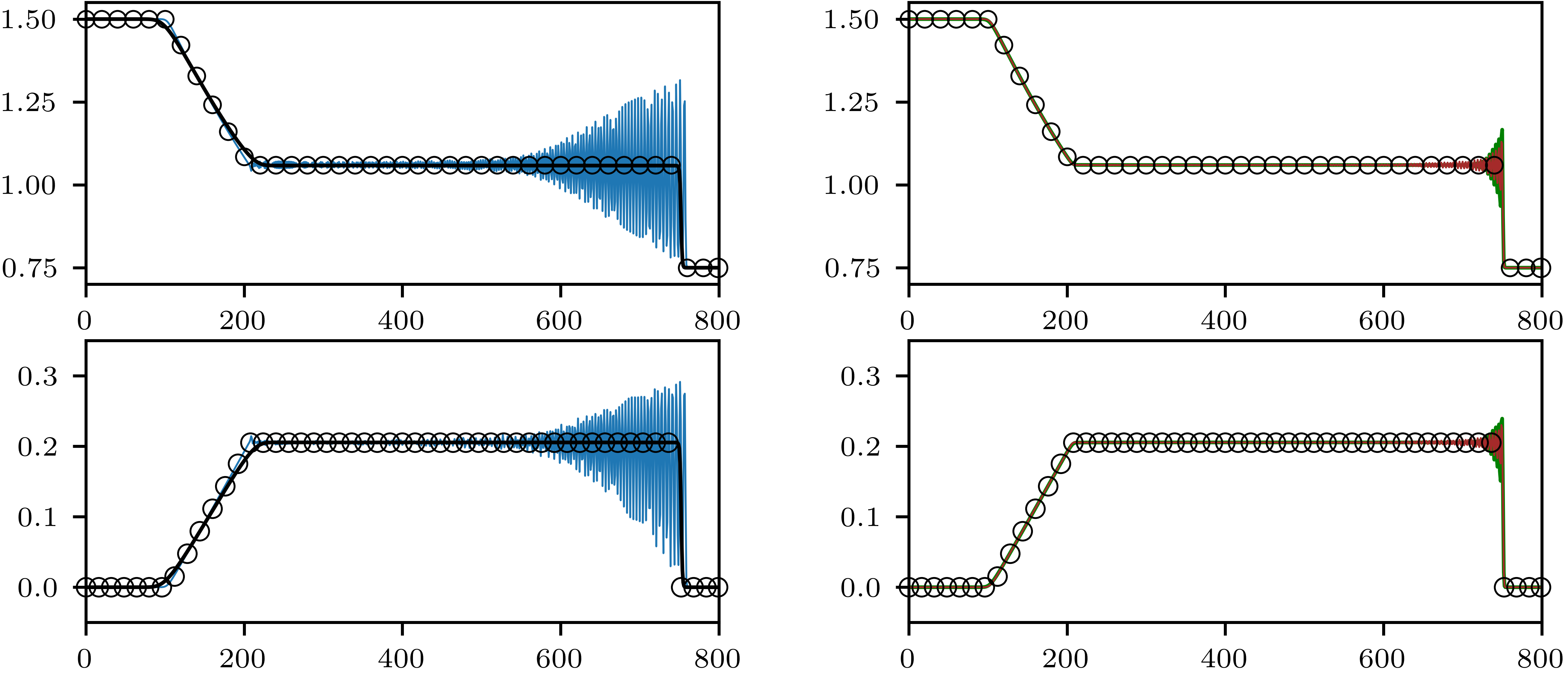}
    \begin{picture}(0,0)
        \put(-401,185){\footnotesize (A)}
        \put(-319.5,-7.5){\footnotesize $x$}
        \put(-435,129){\footnotesize {\rotatebox{90}{Density}}}
        \put(-432,36){\footnotesize {\rotatebox{90}{Velocity}}}
        \put(-180,185){\footnotesize (B)}
        \put(-97.5,-7.5){\footnotesize $x$}
        \put(-212.75,129){\footnotesize {\rotatebox{90}{Density}}}
        \put(-209,36){\footnotesize {\rotatebox{90}{Velocity}}}
    \end{picture}
    \caption{\label{fig:isothermal-shocktube-case1}Isothermal shocktube results obtained using the guided equilibrium on the D2Q9 lattice using different LB schemes at $\theta = 0.35$ and $\nu = 10^{-5}$. Panel (A) shows the OReg (black) and the Lattice-BGK (blue) schemes, while panel (B) shows the first order Essentially Entropic LB (brown) and the projected regularized (green) schemes. The round black symbols correspond to the analytical solution.}
\end{figure*}

\begin{subequations}
	\vspace{-1em}
    \begin{equation}
        u_x = A_0 \sin\left(\frac{2\pi y}{L_y}\right) ,\, u_y = \text{Ma}\sqrt{\theta}
    \end{equation}
    \noindent and that for the rotated wave case is given as:
    \begin{equation}
    \begin{aligned}
            u_x = A_0 \sin&\left(\frac{2\pi}{L_y\sqrt{2}} (-x+y)\right),\\
            u_y = \text{Ma}\sqrt{\theta} + A_0 \sin&\left(\frac{2\pi}{L_y\sqrt{2}} (-x+y)\right).
        \end{aligned}
    \end{equation}
\end{subequations}
\noindent
The wave vectors for the two cases are $\frac{2\pi}{L_y}\hat{j}$ and $\frac{\pi}{L_y} ( -\hat{i} + \hat{j})$ respectively.
The values of the wave amplitude ($A_0$) and spatial discretization ($L_x \times L_y$) are taken to be 0.001 and 1 $\times$ 200 respectively.
Simulations are conducted using the lattice-BGK, OReg and projected regularized (PR) schemes at the reference temperature $\theta = 1/3$ with an imposed kinematic viscosity $\nu = 0.01$ for different Mach numbers.

In the axis-aligned case, the spurious errors due to lattice anisotropy are dormant \citep{PrasianakisKarlinMantzarasBoulouchosPRE2009} and consequently the lattice-BGK and PR schemes recover the correct dissipation rate as shown in Fig.~\ref{fig:wavedecay}(a); it can be seen that the OReg scheme also recovers the correct dissipation rate.
For the rotated wave case, the spurious contributions are activated and, as shown in Fig.~\ref{fig:wavedecay}(b), the uncorrected lattice-BGK and PR models fail to recover the dissipation accurately; the PR scheme yields significantly larger deviations from the expected value as shown on the right vertical axis of Fig.~\ref{fig:wavedecay}(b).
In contrast, the OReg scheme yields the accurate dissipation rate without having to incorporate any correction terms.

The second benchmark we consider is that of a shocktube operating at appreciably small viscosities and lattice temperatures of $\theta \neq 1/3$.
It has been previously shown that the OReg scheme, used with an $\mathcal{O}(u^4)$ polynomial equilibrium representation, exactly recovers the analytical solution in an athermal setting where $\theta = 1/3$ for a lattice viscosity $\nu = 10^{-7}$ \citep{jonnalagaddaPRE2021}.
Here, we examine the behaviour of the OReg scheme with the guided equilibrium when employed at elevated operating temperatures and lattice viscosities of $\theta=0.35, \nu = 10^{-5}$ and $\theta=0.4, \nu = 10^{-9}$, respectively.

\begin{figure*}[t]
\centering
    \centering
    \includegraphics[scale=1]{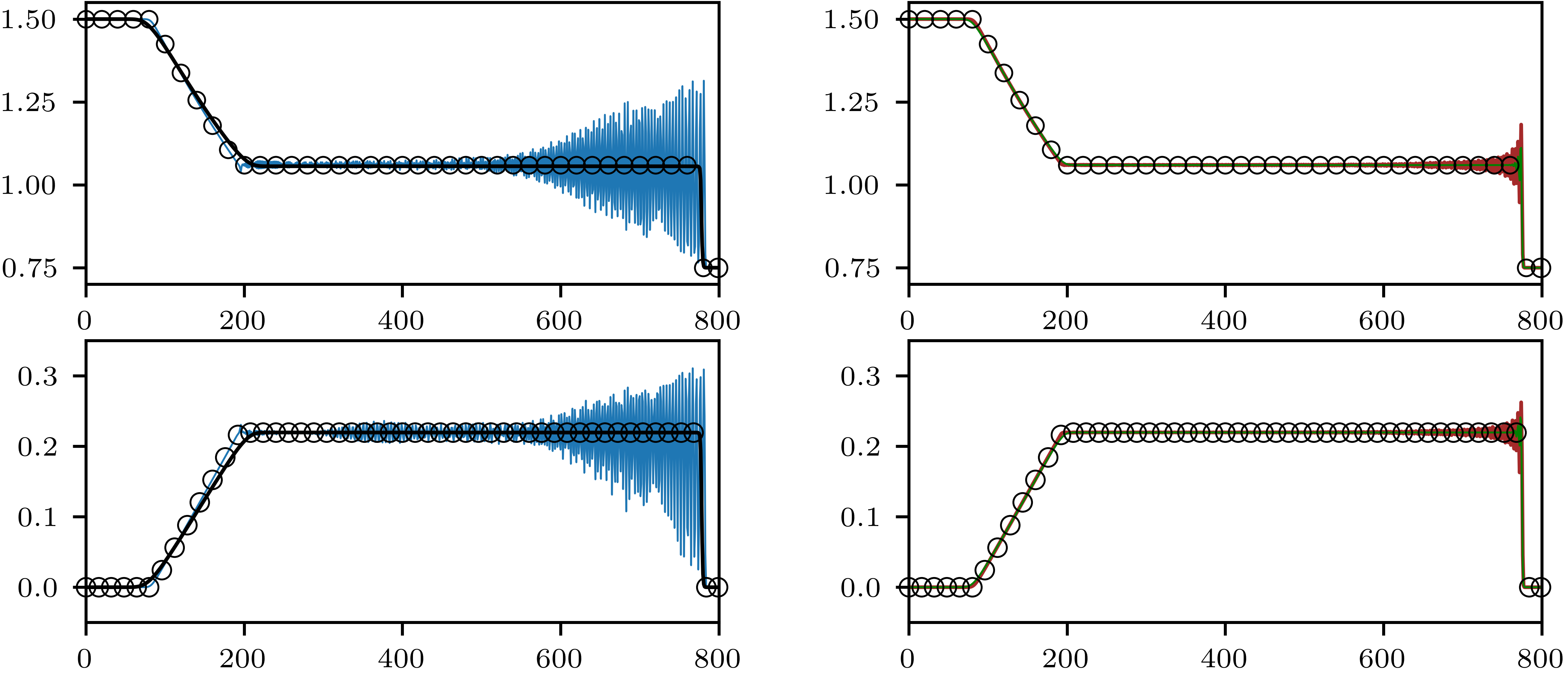}
    \begin{picture}(0,0)
        \put(-401,185){\footnotesize (A)}
        \put(-319.5,-7.5){\footnotesize $x$}
        \put(-435,129){\footnotesize {\rotatebox{90}{Density}}}
        \put(-432,36){\footnotesize {\rotatebox{90}{Velocity}}}

        \put(-180,185){\footnotesize (B)}
        \put(-97.5,-7.5){\footnotesize $x$}
        \put(-212.75,129){\footnotesize {\rotatebox{90}{Density}}}
        \put(-209,36){\footnotesize {\rotatebox{90}{Velocity}}}
    \end{picture}
\vskip1em
\caption{\label{fig:isothermal-shocktube-case2}Isothermal shocktube results obtained using the guided equilibrium on the D2Q9 lattice using different LB schemes at $\theta = 2/5$ and $\nu = 10^{-9}$. The curves have the same meaning as in Fig.~\ref{fig:isothermal-shocktube-case1}.}
\vspace{-1em}
\end{figure*}

In Figs.~(\ref{fig:isothermal-shocktube-case1}) and (\ref{fig:isothermal-shocktube-case2}), we present the numerical results obtained from the lattice-BGK,  the state-of-the-art first-order Essentially Entropic (EE) \citep{atifEELBM}, the projected regularized (PR) and the OReg LB schemes, along with comparisons against analytical solutions.
The simulation is run on a 800$\times$1 grid for 500 time steps and employs the fullway bounce-back treatment for the walls.
It can be seen that while the lattice-BGK yields the largest oscillations and the EE and PR schemes significantly reduces those oscillations, the OReg scheme completely eliminates them.
The results for the EE and PR schemes display an interesting trend, namely, for smaller values of $\theta$ and Re (larger $\nu$), the EE scheme performs relatively better while at higher values of $\theta$ and Re (smaller $\nu$), the PR scheme performs relatively better.
Note that the OReg scheme yields a slightly incorrect slope in the high-density region due to the first-order nature of the resulting hydrodynamics at temperatures of $\theta\neq 1/3$.
However, a computation of the L$_2$ errors reveals that the OReg scheme captures the density with an accuracy of approximately 98.88\% and 98.20\% respectively for the two cases investigated herein.
The simulations conducted on halved and doubled grids also yield similar accuracy hinting at a reasonable level of grid independence of the solution.
Lastly, it is noteworthy that while the accuracy of the EE and PR schemes improves with increasing grid size, it still retains spurious oscillations; in contrast even on halved grids, the OReg scheme demonstrates no spurious oscillations.
\vspace{-1em}

\section{Conclusion}
\vspace{-1em}
\noindent
In conclusion, we have shown that the OReg scheme yields stable and accurate results on computationally efficient first-neighbour lattices, coping with the long-standing limitations induced by the anisotropy error, which have plagued LB simulations so far.
Specifically, with the guided equilibrium defined on the D2Q9 lattice, the OReg scheme is shown to recover isothermal hydrodynamics with third-order accuracy for simulations conducted at the lattice reference temperature and with first-order accuracy at arbitrary lattice temperatures.
Recall that the guided equilibrium is not fully constrained as Eq. (\ref{eq:guidedEq}) does not recover the equilibrium heat flux vector Eq. (\ref{eq:eq-heatflux}); indeed, the deviations in $Q_{\alpha\beta\gamma}'$ and $R_{\alpha\beta\gamma\mu}'$ for a fully constrained equilibrium representation, may yield a more accurate model for isothermal temperatures of $\theta\neq 1/3$.
Nevertheless, even in the current form, the OReg scheme can be coupled to a second population to conduct fully local thermal and compressible flow simulations on standard lattices.
Additionally, in comparison to the bare BGK collision model, the OReg scheme is shown to recover macroscopic dynamics with an accuracy improved by a factor of $\mathcal{O}(u)$ for two equilibrium formulations, further demonstrating the practical usability of the scheme.
Indeed, it can be appreciated that the OReg scheme shifts the focus from modelling accuracy to numerical stability obtained from using different equilibrium formulations.
We also highlight that the OReg scheme presented in Eq. (\ref{eq:Oregf1}) is generic and can be seamlessly generalized to any lattice stencil.
Indeed, the OReg scheme can be directly integrated into high-performance codes such as, e.g., waLBerla \citep{BAUER2021walberla} and highly optimized GPU implementations \citep{montessoriTSLB-pof-2024,lauricellaTSLB2025arxiv}.
Further, the OReg scheme also offers a promising alternative to locally defining the so-called Grad boundary conditions \citep{Latt2008GradBC, DORSCHNER2015GradBC} and coupling to thermal boundary conditions such as those presented in \citep{succi2003ThermalBCs}.
Thus, the OReg scheme presents a significant advance for conducting scalable simulations of physically relevant non-linear dissipative transport problems (see, e.g., \citep{succiPhaseField2005}) characterized by multiscale phenomena,  in complex geometries.
Explorations into these avenues will be the subject of future work.

\vspace{1em}
\begin{acknowledgments}
\vspace{-1em}
\noindent
The authors would like to thank Prof. I.V. Karlin, Dr. Syed Ali Hosseini and the anonymous reviewer for their valuable comments during the preparation of this manuscript.
AA acknowledges DST SERB for the J.C. Bose National Fellowship.
Funded by the European Union - NextGenerationEU.
This work has received funding from the European High Performance Computing Joint Undertaking (JU) (Grant agreement No. 101093169) and the Ministero delle Imprese e del Made in Italy (Grant No. P/040003/01-02/X64).
\end{acknowledgments}

\bibliography{accepted_manuscript}

\begin{thebibliography}{45}%
\makeatletter
\providecommand \@ifxundefined [1]{%
 \@ifx{#1\undefined}
}%
\providecommand \@ifnum [1]{%
 \ifnum #1\expandafter \@firstoftwo
 \else \expandafter \@secondoftwo
 \fi
}%
\providecommand \@ifx [1]{%
 \ifx #1\expandafter \@firstoftwo
 \else \expandafter \@secondoftwo
 \fi
}%
\providecommand \natexlab [1]{#1}%
\providecommand \enquote  [1]{``#1''}%
\providecommand \bibnamefont  [1]{#1}%
\providecommand \bibfnamefont [1]{#1}%
\providecommand \citenamefont [1]{#1}%
\providecommand \href@noop [0]{\@secondoftwo}%
\providecommand \href [0]{\begingroup \@sanitize@url \@href}%
\providecommand \@href[1]{\@@startlink{#1}\@@href}%
\providecommand \@@href[1]{\endgroup#1\@@endlink}%
\providecommand \@sanitize@url [0]{\catcode `\\12\catcode `\$12\catcode
  `\&12\catcode `\#12\catcode `\^12\catcode `\_12\catcode `\%12\relax}%
\providecommand \@@startlink[1]{}%
\providecommand \@@endlink[0]{}%
\providecommand \url  [0]{\begingroup\@sanitize@url \@url }%
\providecommand \@url [1]{\endgroup\@href {#1}{\urlprefix }}%
\providecommand \urlprefix  [0]{URL }%
\providecommand \Eprint [0]{\href }%
\providecommand \doibase [0]{https://doi.org/}%
\providecommand \selectlanguage [0]{\@gobble}%
\providecommand \bibinfo  [0]{\@secondoftwo}%
\providecommand \bibfield  [0]{\@secondoftwo}%
\providecommand \translation [1]{[#1]}%
\providecommand \BibitemOpen [0]{}%
\providecommand \bibitemStop [0]{}%
\providecommand \bibitemNoStop [0]{.\EOS\space}%
\providecommand \EOS [0]{\spacefactor3000\relax}%
\providecommand \BibitemShut  [1]{\csname bibitem#1\endcsname}%
\let\auto@bib@innerbib\@empty
\bibitem [{\citenamefont {Benzi}\ \emph {et~al.}(1992)\citenamefont {Benzi},
  \citenamefont {Succi},\ and\ \citenamefont {Vergassola}}]{benzi1992review}%
  \BibitemOpen
  \bibfield  {author} {\bibinfo {author} {\bibfnamefont {R.}~\bibnamefont
  {Benzi}}, \bibinfo {author} {\bibfnamefont {S.}~\bibnamefont {Succi}},\ and\
  \bibinfo {author} {\bibfnamefont {M.}~\bibnamefont {Vergassola}},\ }\bibfield
   {title} {\bibinfo {title} {The lattice {Boltzmann} equation: theory and
  applications},\ }\href@noop {} {\bibfield  {journal} {\bibinfo  {journal}
  {Physics Reports}\ }\textbf {\bibinfo {volume} {222}},\ \bibinfo {pages}
  {145} (\bibinfo {year} {1992})}\BibitemShut {NoStop}%
\bibitem [{\citenamefont {Chen}\ and\ \citenamefont
  {Doolen}(1998)}]{chen-doolen-ARFM}%
  \BibitemOpen
  \bibfield  {author} {\bibinfo {author} {\bibfnamefont {S.}~\bibnamefont
  {Chen}}\ and\ \bibinfo {author} {\bibfnamefont {G.~D.}\ \bibnamefont
  {Doolen}},\ }\bibfield  {title} {\bibinfo {title} {Lattice {Boltzmann} method
  for fluid flows},\ }\href@noop {} {\bibfield  {journal} {\bibinfo  {journal}
  {Annual review of fluid mechanics}\ }\textbf {\bibinfo {volume} {30}},\
  \bibinfo {pages} {329} (\bibinfo {year} {1998})}\BibitemShut {NoStop}%
\bibitem [{\citenamefont {Aidun}\ and\ \citenamefont
  {Clausen}(2010)}]{aidun2010ARFM}%
  \BibitemOpen
  \bibfield  {author} {\bibinfo {author} {\bibfnamefont {C.~K.}\ \bibnamefont
  {Aidun}}\ and\ \bibinfo {author} {\bibfnamefont {J.~R.}\ \bibnamefont
  {Clausen}},\ }\bibfield  {title} {\bibinfo {title} {Lattice-{Boltzmann}
  method for complex flows},\ }\href@noop {} {\bibfield  {journal} {\bibinfo
  {journal} {Annual review of fluid mechanics}\ }\textbf {\bibinfo {volume}
  {42}},\ \bibinfo {pages} {439} (\bibinfo {year} {2010})}\BibitemShut
  {NoStop}%
\bibitem [{\citenamefont {Tiribocchi}\ \emph {et~al.}(2025)\citenamefont
  {Tiribocchi}, \citenamefont {Durve}, \citenamefont {Lauricella},
  \citenamefont {Montessori}, \citenamefont {Tucny},\ and\ \citenamefont
  {Succi}}]{tiribocciPhysReps2025}%
  \BibitemOpen
  \bibfield  {author} {\bibinfo {author} {\bibfnamefont {A.}~\bibnamefont
  {Tiribocchi}}, \bibinfo {author} {\bibfnamefont {M.}~\bibnamefont {Durve}},
  \bibinfo {author} {\bibfnamefont {M.}~\bibnamefont {Lauricella}}, \bibinfo
  {author} {\bibfnamefont {A.}~\bibnamefont {Montessori}}, \bibinfo {author}
  {\bibfnamefont {J.-M.}\ \bibnamefont {Tucny}},\ and\ \bibinfo {author}
  {\bibfnamefont {S.}~\bibnamefont {Succi}},\ }\bibfield  {title} {\bibinfo
  {title} {Lattice {Boltzmann} simulations for soft flowing matter},\ }\href
  {https://doi.org/https://doi.org/10.1016/j.physrep.2024.11.002} {\bibfield
  {journal} {\bibinfo  {journal} {Physics Reports}\ }\textbf {\bibinfo {volume}
  {1105}},\ \bibinfo {pages} {1} (\bibinfo {year} {2025})}\BibitemShut
  {NoStop}%
\bibitem [{\citenamefont {Succi}(2001)}]{succi2001lattice}%
  \BibitemOpen
  \bibfield  {author} {\bibinfo {author} {\bibfnamefont {S.}~\bibnamefont
  {Succi}},\ }\href@noop {} {\emph {\bibinfo {title} {{The lattice Boltzmann
  equation: for fluid dynamics and beyond}}}}\ (\bibinfo  {publisher} {Oxford
  University Press},\ \bibinfo {year} {2001})\BibitemShut {NoStop}%
\bibitem [{\citenamefont {{Kr{\"u}ger}}\ \emph {et~al.}(2017)\citenamefont
  {{Kr{\"u}ger}}, \citenamefont {{Halim Kusumaatmaja}}, \citenamefont
  {{Alexandr Kuzmin}}, \citenamefont {{Orest Shardt}}, \citenamefont {{Goncalo
  Silva}},\ and\ \citenamefont {{Erlend Mahnus Viggen}}}]{kruger}%
  \BibitemOpen
  \bibfield  {author} {\bibinfo {author} {\bibfnamefont {T.}~\bibnamefont
  {{Kr{\"u}ger}}}, \bibinfo {author} {\bibnamefont {{Halim Kusumaatmaja}}},
  \bibinfo {author} {\bibnamefont {{Alexandr Kuzmin}}}, \bibinfo {author}
  {\bibnamefont {{Orest Shardt}}}, \bibinfo {author} {\bibnamefont {{Goncalo
  Silva}}},\ and\ \bibinfo {author} {\bibnamefont {{Erlend Mahnus Viggen}}},\
  }\href@noop {} {\emph {\bibinfo {title} {The Lattice Boltzmann Method:
  Principles and Practice}}},\ \bibinfo {edition} {1st}\ ed.\ (\bibinfo
  {publisher} {Springer International Publishing},\ \bibinfo {year}
  {2017})\BibitemShut {NoStop}%
\bibitem [{\citenamefont {Succi}(2018)}]{succi2018lattice}%
  \BibitemOpen
  \bibfield  {author} {\bibinfo {author} {\bibfnamefont {S.}~\bibnamefont
  {Succi}},\ }\href@noop {} {\emph {\bibinfo {title} {{The lattice Boltzmann
  equation: for complex states of flowing matter}}}}\ (\bibinfo  {publisher}
  {Oxford University Press},\ \bibinfo {year} {2018})\BibitemShut {NoStop}%
\bibitem [{\citenamefont {Jonnalagadda}\ \emph {et~al.}(2023)\citenamefont
  {Jonnalagadda}, \citenamefont {Sharma},\ and\ \citenamefont
  {Agrawal}}]{jonnalagadda2023NHTB}%
  \BibitemOpen
  \bibfield  {author} {\bibinfo {author} {\bibfnamefont {A.}~\bibnamefont
  {Jonnalagadda}}, \bibinfo {author} {\bibfnamefont {A.}~\bibnamefont
  {Sharma}},\ and\ \bibinfo {author} {\bibfnamefont {A.}~\bibnamefont
  {Agrawal}},\ }\bibfield  {title} {\bibinfo {title} {{On application of the
  regularized lattice Boltzmann method for isothermal flows with non-vanishing
  Knudsen numbers}},\ }\href {https://doi.org/10.1080/10407790.2023.2220908}
  {\bibfield  {journal} {\bibinfo  {journal} {Numerical Heat Transfer, Part B:
  Fundamentals}\ }\textbf {\bibinfo {volume} {84}},\ \bibinfo {pages} {756}
  (\bibinfo {year} {2023})}\BibitemShut {NoStop}%
\bibitem [{\citenamefont {Agrawal}\ \emph {et~al.}(2020)\citenamefont
  {Agrawal}, \citenamefont {Kushwaha},\ and\ \citenamefont
  {Jadhav}}]{amitagrawalbook}%
  \BibitemOpen
  \bibfield  {author} {\bibinfo {author} {\bibfnamefont {A.}~\bibnamefont
  {Agrawal}}, \bibinfo {author} {\bibfnamefont {H.~M.}\ \bibnamefont
  {Kushwaha}},\ and\ \bibinfo {author} {\bibfnamefont {R.~S.}\ \bibnamefont
  {Jadhav}},\ }\href {https://doi.org/10.1007/978-3-030-10662-1} {\emph
  {\bibinfo {title} {Microscale flow and heat transfer}}}\ (\bibinfo
  {publisher} {Springer},\ \bibinfo {year} {2020})\BibitemShut {NoStop}%
\bibitem [{\citenamefont {Singh}\ and\ \citenamefont {Agrawal}(2016)}]{o13}%
  \BibitemOpen
  \bibfield  {author} {\bibinfo {author} {\bibfnamefont {N.}~\bibnamefont
  {Singh}}\ and\ \bibinfo {author} {\bibfnamefont {A.}~\bibnamefont
  {Agrawal}},\ }\bibfield  {title} {\bibinfo {title}
  {{Onsager's-principle-consistent 13-moment transport equations}},\ }\bibfield
   {journal} {\bibinfo  {journal} {Phys. Rev. E}\ }\textbf {\bibinfo {volume}
  {93}},\ \href {https://doi.org/10.1103/PhysRevE.93.063111}
  {10.1103/PhysRevE.93.063111} (\bibinfo {year} {2016})\BibitemShut {NoStop}%
\bibitem [{\citenamefont {Singh}\ \emph {et~al.}(2017)\citenamefont {Singh},
  \citenamefont {Jadhav},\ and\ \citenamefont {Agrawal}}]{oBurnett}%
  \BibitemOpen
  \bibfield  {author} {\bibinfo {author} {\bibfnamefont {N.}~\bibnamefont
  {Singh}}, \bibinfo {author} {\bibfnamefont {R.~S.}\ \bibnamefont {Jadhav}},\
  and\ \bibinfo {author} {\bibfnamefont {A.}~\bibnamefont {Agrawal}},\
  }\bibfield  {title} {\bibinfo {title} {{Derivation of stable Burnett
  equations for rarefied gas flows}},\ }\bibfield  {journal} {\bibinfo
  {journal} {Phys. Rev. E}\ }\textbf {\bibinfo {volume} {96}},\ \href
  {https://doi.org/10.1103/PhysRevE.96.013106} {10.1103/PhysRevE.96.013106}
  (\bibinfo {year} {2017})\BibitemShut {NoStop}%
\bibitem [{\citenamefont {Yadav}\ \emph {et~al.}(2023)\citenamefont {Yadav},
  \citenamefont {Jonnalagadda},\ and\ \citenamefont
  {Agrawal}}]{upendraAIP2023}%
  \BibitemOpen
  \bibfield  {author} {\bibinfo {author} {\bibfnamefont {U.}~\bibnamefont
  {Yadav}}, \bibinfo {author} {\bibfnamefont {A.}~\bibnamefont
  {Jonnalagadda}},\ and\ \bibinfo {author} {\bibfnamefont {A.}~\bibnamefont
  {Agrawal}},\ }\bibfield  {title} {\bibinfo {title} {Third-order accurate
  13-moment equations for non-continuum transport phenomenon},\ }\href
  {https://doi.org/10.1063/5.0143420} {\bibfield  {journal} {\bibinfo
  {journal} {AIP Advances}\ }\textbf {\bibinfo {volume} {13}},\ \bibinfo
  {pages} {045311} (\bibinfo {year} {2023})}\BibitemShut {NoStop}%
\bibitem [{\citenamefont {Yadav}\ \emph {et~al.}(2024)\citenamefont {Yadav},
  \citenamefont {Jonnalagadda},\ and\ \citenamefont
  {Agrawal}}]{upendraJFM2023}%
  \BibitemOpen
  \bibfield  {author} {\bibinfo {author} {\bibfnamefont {U.}~\bibnamefont
  {Yadav}}, \bibinfo {author} {\bibfnamefont {A.}~\bibnamefont
  {Jonnalagadda}},\ and\ \bibinfo {author} {\bibfnamefont {A.}~\bibnamefont
  {Agrawal}},\ }\bibfield  {title} {\bibinfo {title} {{Derivation of
  extended-OBurnett and super-OBurnett equations and their analytical solution
  to plane Poiseuille flow at non-zero Knudsen number}},\ }\href
  {https://doi.org/10.1017/jfm.2024.141} {\bibfield  {journal} {\bibinfo
  {journal} {Journal of Fluid Mechanics}\ }\textbf {\bibinfo {volume} {983}},\
  \bibinfo {pages} {A29} (\bibinfo {year} {2024})}\BibitemShut {NoStop}%
\bibitem [{\citenamefont {Chen}\ \emph {et~al.}(1994)\citenamefont {Chen},
  \citenamefont {Ohashi},\ and\ \citenamefont
  {Akiyama}}]{ChenOhashiAkiyamaPRE1994}%
  \BibitemOpen
  \bibfield  {author} {\bibinfo {author} {\bibfnamefont {Y.}~\bibnamefont
  {Chen}}, \bibinfo {author} {\bibfnamefont {H.}~\bibnamefont {Ohashi}},\ and\
  \bibinfo {author} {\bibfnamefont {M.}~\bibnamefont {Akiyama}},\ }\bibfield
  {title} {\bibinfo {title} {{Thermal lattice Bhatnagar-Gross-Krook model
  without nonlinear deviations in macrodynamic equations}},\ }\href
  {https://doi.org/10.1103/PhysRevE.50.2776} {\bibfield  {journal} {\bibinfo
  {journal} {Phys. Rev. E}\ }\textbf {\bibinfo {volume} {50}},\ \bibinfo
  {pages} {2776} (\bibinfo {year} {1994})}\BibitemShut {NoStop}%
\bibitem [{\citenamefont {Wagner}\ and\ \citenamefont
  {Yeomans}(1999)}]{WagnerYeomansPRE1999}%
  \BibitemOpen
  \bibfield  {author} {\bibinfo {author} {\bibfnamefont {A.~J.}\ \bibnamefont
  {Wagner}}\ and\ \bibinfo {author} {\bibfnamefont {J.~M.}\ \bibnamefont
  {Yeomans}},\ }\bibfield  {title} {\bibinfo {title} {{Phase separation under
  shear in two-dimensional binary fluids}},\ }\href
  {https://doi.org/10.1103/PhysRevE.59.4366} {\bibfield  {journal} {\bibinfo
  {journal} {Phys. Rev. E}\ }\textbf {\bibinfo {volume} {59}},\ \bibinfo
  {pages} {4366} (\bibinfo {year} {1999})}\BibitemShut {NoStop}%
\bibitem [{\citenamefont {Dellar}(2002)}]{dellarPRE2002}%
  \BibitemOpen
  \bibfield  {author} {\bibinfo {author} {\bibfnamefont {P.~J.}\ \bibnamefont
  {Dellar}},\ }\bibfield  {title} {\bibinfo {title} {{Nonhydrodynamic modes and
  a priori construction of shallow water lattice Boltzmann equations}},\ }\href
  {https://doi.org/10.1103/PhysRevE.65.036309} {\bibfield  {journal} {\bibinfo
  {journal} {Phys. Rev. E}\ }\textbf {\bibinfo {volume} {65}},\ \bibinfo
  {pages} {036309} (\bibinfo {year} {2002})}\BibitemShut {NoStop}%
\bibitem [{\citenamefont {Dellar}(2003)}]{dellarJCP2003}%
  \BibitemOpen
  \bibfield  {author} {\bibinfo {author} {\bibfnamefont {P.~J.}\ \bibnamefont
  {Dellar}},\ }\bibfield  {title} {\bibinfo {title} {{Incompressible limits of
  lattice Boltzmann equations using multiple relaxation times}},\ }\href
  {https://doi.org/https://doi.org/10.1016/S0021-9991(03)00279-1} {\bibfield
  {journal} {\bibinfo  {journal} {Journal of Computational Physics}\ }\textbf
  {\bibinfo {volume} {190}},\ \bibinfo {pages} {351} (\bibinfo {year}
  {2003})}\BibitemShut {NoStop}%
\bibitem [{\citenamefont {Házi}\ and\ \citenamefont
  {Kávrán}(2006)}]{HaziKavranJPA2006}%
  \BibitemOpen
  \bibfield  {author} {\bibinfo {author} {\bibfnamefont {G.}~\bibnamefont
  {Házi}}\ and\ \bibinfo {author} {\bibfnamefont {P.}~\bibnamefont
  {Kávrán}},\ }\bibfield  {title} {\bibinfo {title} {{On the cubic velocity
  deviations in lattice Boltzmann methods}},\ }\href
  {https://doi.org/10.1088/0305-4470/39/12/019} {\bibfield  {journal} {\bibinfo
   {journal} {Journal of Physics A: Mathematical and General}\ }\textbf
  {\bibinfo {volume} {39}},\ \bibinfo {pages} {3127} (\bibinfo {year}
  {2006})}\BibitemShut {NoStop}%
\bibitem [{\citenamefont {Házi}\ and\ \citenamefont
  {Jiménez}(2006)}]{HaziJimenezCompFlu2006}%
  \BibitemOpen
  \bibfield  {author} {\bibinfo {author} {\bibfnamefont {G.}~\bibnamefont
  {Házi}}\ and\ \bibinfo {author} {\bibfnamefont {C.}~\bibnamefont
  {Jiménez}},\ }\bibfield  {title} {\bibinfo {title} {{Simulation of
  two-dimensional decaying turbulence using the “incompressible” extensions
  of the lattice Boltzmann method}},\ }\href
  {https://doi.org/https://doi.org/10.1016/j.compfluid.2004.12.003} {\bibfield
  {journal} {\bibinfo  {journal} {Computers \& Fluids}\ }\textbf {\bibinfo
  {volume} {35}},\ \bibinfo {pages} {280} (\bibinfo {year} {2006})}\BibitemShut
  {NoStop}%
\bibitem [{\citenamefont {Keating}\ \emph {et~al.}(2007)\citenamefont
  {Keating}, \citenamefont {Vahala}, \citenamefont {Yepez}, \citenamefont
  {Soe},\ and\ \citenamefont {Vahala}}]{KeatingValhalaYepezSoeValhalaPRE2007}%
  \BibitemOpen
  \bibfield  {author} {\bibinfo {author} {\bibfnamefont {B.}~\bibnamefont
  {Keating}}, \bibinfo {author} {\bibfnamefont {G.}~\bibnamefont {Vahala}},
  \bibinfo {author} {\bibfnamefont {J.}~\bibnamefont {Yepez}}, \bibinfo
  {author} {\bibfnamefont {M.}~\bibnamefont {Soe}},\ and\ \bibinfo {author}
  {\bibfnamefont {L.}~\bibnamefont {Vahala}},\ }\bibfield  {title} {\bibinfo
  {title} {{Entropic lattice Boltzmann representations required to recover
  Navier-Stokes flows}},\ }\href {https://doi.org/10.1103/PhysRevE.75.036712}
  {\bibfield  {journal} {\bibinfo  {journal} {Phys. Rev. E}\ }\textbf {\bibinfo
  {volume} {75}},\ \bibinfo {pages} {036712} (\bibinfo {year}
  {2007})}\BibitemShut {NoStop}%
\bibitem [{\citenamefont {Dellar}(2014)}]{dellarJCP2014}%
  \BibitemOpen
  \bibfield  {author} {\bibinfo {author} {\bibfnamefont {P.~J.}\ \bibnamefont
  {Dellar}},\ }\bibfield  {title} {\bibinfo {title} {{Lattice Boltzmann
  algorithms without cubic defects in Galilean invariance on standard
  lattices}},\ }\href
  {https://doi.org/https://doi.org/10.1016/j.jcp.2013.11.021} {\bibfield
  {journal} {\bibinfo  {journal} {Journal of Computational Physics}\ }\textbf
  {\bibinfo {volume} {259}},\ \bibinfo {pages} {270} (\bibinfo {year}
  {2014})}\BibitemShut {NoStop}%
\bibitem [{\citenamefont {Prasianakis}\ and\ \citenamefont
  {Karlin}(2007)}]{PrasianakisKarlin2007}%
  \BibitemOpen
  \bibfield  {author} {\bibinfo {author} {\bibfnamefont {N.~I.}\ \bibnamefont
  {Prasianakis}}\ and\ \bibinfo {author} {\bibfnamefont {I.~V.}\ \bibnamefont
  {Karlin}},\ }\bibfield  {title} {\bibinfo {title} {{Lattice Boltzmann method
  for thermal flow simulation on standard lattices}},\ }\href
  {https://doi.org/10.1103/PhysRevE.76.016702} {\bibfield  {journal} {\bibinfo
  {journal} {Phys. Rev. E}\ }\textbf {\bibinfo {volume} {76}},\ \bibinfo
  {pages} {016702} (\bibinfo {year} {2007})}\BibitemShut {NoStop}%
\bibitem [{\citenamefont {Prasianakis}\ and\ \citenamefont
  {Karlin}(2008)}]{PrasianakisKarlin2008}%
  \BibitemOpen
  \bibfield  {author} {\bibinfo {author} {\bibfnamefont {N.~I.}\ \bibnamefont
  {Prasianakis}}\ and\ \bibinfo {author} {\bibfnamefont {I.~V.}\ \bibnamefont
  {Karlin}},\ }\bibfield  {title} {\bibinfo {title} {{Lattice Boltzmann method
  for simulation of compressible flows on standard lattices}},\ }\href
  {https://doi.org/10.1103/PhysRevE.78.016704} {\bibfield  {journal} {\bibinfo
  {journal} {Phys. Rev. E}\ }\textbf {\bibinfo {volume} {78}},\ \bibinfo
  {pages} {016704} (\bibinfo {year} {2008})}\BibitemShut {NoStop}%
\bibitem [{\citenamefont {Prasianakis}\ \emph {et~al.}(2009)\citenamefont
  {Prasianakis}, \citenamefont {Karlin}, \citenamefont {Mantzaras},\ and\
  \citenamefont {Boulouchos}}]{PrasianakisKarlinMantzarasBoulouchosPRE2009}%
  \BibitemOpen
  \bibfield  {author} {\bibinfo {author} {\bibfnamefont {N.~I.}\ \bibnamefont
  {Prasianakis}}, \bibinfo {author} {\bibfnamefont {I.~V.}\ \bibnamefont
  {Karlin}}, \bibinfo {author} {\bibfnamefont {J.}~\bibnamefont {Mantzaras}},\
  and\ \bibinfo {author} {\bibfnamefont {K.~B.}\ \bibnamefont {Boulouchos}},\
  }\bibfield  {title} {\bibinfo {title} {{Lattice Boltzmann method with
  restored Galilean invariance}},\ }\href
  {https://doi.org/10.1103/PhysRevE.79.066702} {\bibfield  {journal} {\bibinfo
  {journal} {Phys. Rev. E}\ }\textbf {\bibinfo {volume} {79}},\ \bibinfo
  {pages} {066702} (\bibinfo {year} {2009})}\BibitemShut {NoStop}%
\bibitem [{\citenamefont {Saadat}\ \emph {et~al.}(2019)\citenamefont {Saadat},
  \citenamefont {B\"osch},\ and\ \citenamefont
  {Karlin}}]{SaadatBoschKarlinPRE2019}%
  \BibitemOpen
  \bibfield  {author} {\bibinfo {author} {\bibfnamefont {M.~H.}\ \bibnamefont
  {Saadat}}, \bibinfo {author} {\bibfnamefont {F.}~\bibnamefont {B\"osch}},\
  and\ \bibinfo {author} {\bibfnamefont {I.~V.}\ \bibnamefont {Karlin}},\
  }\bibfield  {title} {\bibinfo {title} {{Lattice Boltzmann model for
  compressible flows on standard lattices: Variable Prandtl number and
  adiabatic exponent}},\ }\href {https://doi.org/10.1103/PhysRevE.99.013306}
  {\bibfield  {journal} {\bibinfo  {journal} {Phys. Rev. E}\ }\textbf {\bibinfo
  {volume} {99}},\ \bibinfo {pages} {013306} (\bibinfo {year}
  {2019})}\BibitemShut {NoStop}%
\bibitem [{\citenamefont {Saadat}\ \emph {et~al.}(2021)\citenamefont {Saadat},
  \citenamefont {Dorschner},\ and\ \citenamefont
  {Karlin}}]{SaadatDorschnerKarlin2021}%
  \BibitemOpen
  \bibfield  {author} {\bibinfo {author} {\bibfnamefont {M.~H.}\ \bibnamefont
  {Saadat}}, \bibinfo {author} {\bibfnamefont {B.}~\bibnamefont {Dorschner}},\
  and\ \bibinfo {author} {\bibfnamefont {I.}~\bibnamefont {Karlin}},\
  }\bibfield  {title} {\bibinfo {title} {{Extended Lattice Boltzmann Model}},\
  }\bibfield  {journal} {\bibinfo  {journal} {Entropy}\ }\textbf {\bibinfo
  {volume} {23}},\ \href {https://doi.org/10.3390/e23040475}
  {10.3390/e23040475} (\bibinfo {year} {2021})\BibitemShut {NoStop}%
\bibitem [{\citenamefont {Jonnalagadda}\ \emph
  {et~al.}(2021{\natexlab{a}})\citenamefont {Jonnalagadda}, \citenamefont
  {Sharma},\ and\ \citenamefont {Agrawal}}]{jonnalagaddaJHT2021}%
  \BibitemOpen
  \bibfield  {author} {\bibinfo {author} {\bibfnamefont {A.}~\bibnamefont
  {Jonnalagadda}}, \bibinfo {author} {\bibfnamefont {A.}~\bibnamefont
  {Sharma}},\ and\ \bibinfo {author} {\bibfnamefont {A.}~\bibnamefont
  {Agrawal}},\ }\bibfield  {title} {\bibinfo {title} {{Revisiting the Lattice
  Boltzmann Method Through a Nonequilibrium Thermodynamics Perspective}},\
  }\href@noop {} {\bibfield  {journal} {\bibinfo  {journal} {Journal of Heat
  Transfer}\ }\textbf {\bibinfo {volume} {143}} (\bibinfo {year}
  {2021}{\natexlab{a}})},\ \bibinfo {note} {052102}\BibitemShut {NoStop}%
\bibitem [{\citenamefont {Jonnalagadda}\ \emph
  {et~al.}(2021{\natexlab{b}})\citenamefont {Jonnalagadda}, \citenamefont
  {Sharma},\ and\ \citenamefont {Agrawal}}]{jonnalagaddaPRE2021}%
  \BibitemOpen
  \bibfield  {author} {\bibinfo {author} {\bibfnamefont {A.}~\bibnamefont
  {Jonnalagadda}}, \bibinfo {author} {\bibfnamefont {A.}~\bibnamefont
  {Sharma}},\ and\ \bibinfo {author} {\bibfnamefont {A.}~\bibnamefont
  {Agrawal}},\ }\bibfield  {title} {\bibinfo {title} {{Onsager-regularized
  lattice Boltzmann method: A nonequilibrium thermodynamics-based regularized
  lattice Boltzmann method}},\ }\href@noop {} {\bibfield  {journal} {\bibinfo
  {journal} {Phys. Rev. E}\ }\textbf {\bibinfo {volume} {104}},\ \bibinfo
  {pages} {015313} (\bibinfo {year} {2021}{\natexlab{b}})}\BibitemShut
  {NoStop}%
\bibitem [{\citenamefont {Mahendra}(2011)}]{mahendra-thesis}%
  \BibitemOpen
  \bibfield  {author} {\bibinfo {author} {\bibfnamefont {A.~K.}\ \bibnamefont
  {Mahendra}},\ }\emph {\bibinfo {title} {{Meshless method for Slip flows}}},\
  \href@noop {} {Ph.D. thesis},\ \bibinfo  {school} {Homi Bhabha National
  Institute} (\bibinfo {year} {2011})\BibitemShut {NoStop}%
\bibitem [{\citenamefont {Mahendra}\ and\ \citenamefont
  {Singh}(2013)}]{mahendra2013onsager}%
  \BibitemOpen
  \bibfield  {author} {\bibinfo {author} {\bibfnamefont {A.~K.}\ \bibnamefont
  {Mahendra}}\ and\ \bibinfo {author} {\bibfnamefont {R.~K.}\ \bibnamefont
  {Singh}},\ }\href {https://arxiv.org/abs/1308.4119} {\bibinfo {title}
  {{Onsager reciprocity principle for kinetic models and kinetic schemes}}}
  (\bibinfo {year} {2013}),\ \Eprint {https://arxiv.org/abs/1308.4119}
  {arXiv:1308.4119 [physics.flu-dyn]} \BibitemShut {NoStop}%
\bibitem [{\citenamefont {Hosseini}\ \emph {et~al.}(2023)\citenamefont
  {Hosseini}, \citenamefont {Atif}, \citenamefont {Ansumali},\ and\
  \citenamefont {Karlin}}]{hosseini2023ELBMreview}%
  \BibitemOpen
  \bibfield  {author} {\bibinfo {author} {\bibfnamefont {S.}~\bibnamefont
  {Hosseini}}, \bibinfo {author} {\bibfnamefont {M.}~\bibnamefont {Atif}},
  \bibinfo {author} {\bibfnamefont {S.}~\bibnamefont {Ansumali}},\ and\
  \bibinfo {author} {\bibfnamefont {I.}~\bibnamefont {Karlin}},\ }\bibfield
  {title} {\bibinfo {title} {{Entropic lattice Boltzmann methods: A review}},\
  }\href {https://doi.org/https://doi.org/10.1016/j.compfluid.2023.105884}
  {\bibfield  {journal} {\bibinfo  {journal} {Computers \& Fluids}\ }\textbf
  {\bibinfo {volume} {259}},\ \bibinfo {pages} {105884} (\bibinfo {year}
  {2023})}\BibitemShut {NoStop}%
\bibitem [{\citenamefont {Latt}\ and\ \citenamefont
  {Chopard}(2006)}]{LATT2006}%
  \BibitemOpen
  \bibfield  {author} {\bibinfo {author} {\bibfnamefont {J.}~\bibnamefont
  {Latt}}\ and\ \bibinfo {author} {\bibfnamefont {B.}~\bibnamefont {Chopard}},\
  }\bibfield  {title} {\bibinfo {title} {{Lattice Boltzmann method with
  regularized pre-collision distribution functions}},\ }\bibfield  {journal}
  {\bibinfo  {journal} {Mathematics and Computers in Simulation}\ }\textbf
  {\bibinfo {volume} {72}},\ \href
  {https://doi.org/10.1016/j.matcom.2006.05.017} {10.1016/j.matcom.2006.05.017}
  (\bibinfo {year} {2006})\BibitemShut {NoStop}%
\bibitem [{\citenamefont {Chen}\ \emph {et~al.}(2020)\citenamefont {Chen},
  \citenamefont {Zhang},\ and\ \citenamefont
  {Gopalakrishnan}}]{chen2020filtered}%
  \BibitemOpen
  \bibfield  {author} {\bibinfo {author} {\bibfnamefont {H.}~\bibnamefont
  {Chen}}, \bibinfo {author} {\bibfnamefont {R.}~\bibnamefont {Zhang}},\ and\
  \bibinfo {author} {\bibfnamefont {P.}~\bibnamefont {Gopalakrishnan}},\
  }\bibfield  {title} {\bibinfo {title} {{Filtered lattice Boltzmann collision
  formulation enforcing isotropy and Galilean invariance}},\ }\href@noop {}
  {\bibfield  {journal} {\bibinfo  {journal} {Physica Scripta}\ }\textbf
  {\bibinfo {volume} {95}},\ \bibinfo {pages} {034003} (\bibinfo {year}
  {2020})}\BibitemShut {NoStop}%
\bibitem [{\citenamefont {Malaspinas}(2015)}]{malaspinasArXiV}%
  \BibitemOpen
  \bibfield  {author} {\bibinfo {author} {\bibfnamefont {O.}~\bibnamefont
  {Malaspinas}},\ }\href {https://arxiv.org/abs/1505.06900} {\bibinfo {title}
  {{Increasing stability and accuracy of the lattice Boltzmann scheme:
  recursivity and regularization}}} (\bibinfo {year} {2015}),\ \Eprint
  {https://arxiv.org/abs/1505.06900} {arXiv:1505.06900 [physics.flu-dyn]}
  \BibitemShut {NoStop}%
\bibitem [{\citenamefont {Coreixas}\ \emph {et~al.}(2017)\citenamefont
  {Coreixas}, \citenamefont {Wissocq}, \citenamefont {Puigt}, \citenamefont
  {Boussuge},\ and\ \citenamefont {Sagaut}}]{coreixasRecReg}%
  \BibitemOpen
  \bibfield  {author} {\bibinfo {author} {\bibfnamefont {C.}~\bibnamefont
  {Coreixas}}, \bibinfo {author} {\bibfnamefont {G.}~\bibnamefont {Wissocq}},
  \bibinfo {author} {\bibfnamefont {G.}~\bibnamefont {Puigt}}, \bibinfo
  {author} {\bibfnamefont {J.-F.}\ \bibnamefont {Boussuge}},\ and\ \bibinfo
  {author} {\bibfnamefont {P.}~\bibnamefont {Sagaut}},\ }\bibfield  {title}
  {\bibinfo {title} {{Recursive regularization step for high-order lattice
  Boltzmann methods}},\ }\bibfield  {journal} {\bibinfo  {journal} {Phys. Rev.
  E}\ }\textbf {\bibinfo {volume} {96}},\ \href
  {https://doi.org/10.1103/PhysRevE.96.033306} {10.1103/PhysRevE.96.033306}
  (\bibinfo {year} {2017})\BibitemShut {NoStop}%
\bibitem [{\citenamefont {Feng}\ \emph {et~al.}(2019)\citenamefont {Feng},
  \citenamefont {Boivin}, \citenamefont {Jacob},\ and\ \citenamefont
  {Sagaut}}]{fengHybridRecreg2019}%
  \BibitemOpen
  \bibfield  {author} {\bibinfo {author} {\bibfnamefont {Y.}~\bibnamefont
  {Feng}}, \bibinfo {author} {\bibfnamefont {P.}~\bibnamefont {Boivin}},
  \bibinfo {author} {\bibfnamefont {J.}~\bibnamefont {Jacob}},\ and\ \bibinfo
  {author} {\bibfnamefont {P.}~\bibnamefont {Sagaut}},\ }\bibfield  {title}
  {\bibinfo {title} {{Hybrid recursive regularized thermal lattice Boltzmann
  model for high subsonic compressible flows}},\ }\href
  {https://doi.org/https://doi.org/10.1016/j.jcp.2019.05.031} {\bibfield
  {journal} {\bibinfo  {journal} {Journal of Computational Physics}\ }\textbf
  {\bibinfo {volume} {394}},\ \bibinfo {pages} {82} (\bibinfo {year}
  {2019})}\BibitemShut {NoStop}%
\bibitem [{\citenamefont {Ansumali}\ and\ \citenamefont
  {Karlin}(2005)}]{consistentLBM}%
  \BibitemOpen
  \bibfield  {author} {\bibinfo {author} {\bibfnamefont {S.}~\bibnamefont
  {Ansumali}}\ and\ \bibinfo {author} {\bibfnamefont {I.~V.}\ \bibnamefont
  {Karlin}},\ }\bibfield  {title} {\bibinfo {title} {{Consistent Lattice
  Boltzmann Method}},\ }\href {https://doi.org/10.1103/PhysRevLett.95.260605}
  {\bibfield  {journal} {\bibinfo  {journal} {Phys. Rev. Lett.}\ }\textbf
  {\bibinfo {volume} {95}},\ \bibinfo {pages} {260605} (\bibinfo {year}
  {2005})}\BibitemShut {NoStop}%
\bibitem [{\citenamefont {Atif}\ \emph {et~al.}(2017)\citenamefont {Atif},
  \citenamefont {Kolluru}, \citenamefont {Thantanapally},\ and\ \citenamefont
  {Ansumali}}]{atifEELBM}%
  \BibitemOpen
  \bibfield  {author} {\bibinfo {author} {\bibfnamefont {M.}~\bibnamefont
  {Atif}}, \bibinfo {author} {\bibfnamefont {P.~K.}\ \bibnamefont {Kolluru}},
  \bibinfo {author} {\bibfnamefont {C.}~\bibnamefont {Thantanapally}},\ and\
  \bibinfo {author} {\bibfnamefont {S.}~\bibnamefont {Ansumali}},\ }\bibfield
  {title} {\bibinfo {title} {{Essentially Entropic Lattice Boltzmann Model}},\
  }\href {https://doi.org/10.1103/PhysRevLett.119.240602} {\bibfield  {journal}
  {\bibinfo  {journal} {Phys. Rev. Lett.}\ }\textbf {\bibinfo {volume} {119}},\
  \bibinfo {pages} {240602} (\bibinfo {year} {2017})}\BibitemShut {NoStop}%
\bibitem [{\citenamefont {Bauer}\ \emph {et~al.}(2021)\citenamefont {Bauer},
  \citenamefont {Eibl}, \citenamefont {Godenschwager}, \citenamefont {Kohl},
  \citenamefont {Kuron}, \citenamefont {Rettinger}, \citenamefont {Schornbaum},
  \citenamefont {Schwarzmeier}, \citenamefont {Thönnes}, \citenamefont
  {Köstler},\ and\ \citenamefont {Rüde}}]{BAUER2021walberla}%
  \BibitemOpen
  \bibfield  {author} {\bibinfo {author} {\bibfnamefont {M.}~\bibnamefont
  {Bauer}}, \bibinfo {author} {\bibfnamefont {S.}~\bibnamefont {Eibl}},
  \bibinfo {author} {\bibfnamefont {C.}~\bibnamefont {Godenschwager}}, \bibinfo
  {author} {\bibfnamefont {N.}~\bibnamefont {Kohl}}, \bibinfo {author}
  {\bibfnamefont {M.}~\bibnamefont {Kuron}}, \bibinfo {author} {\bibfnamefont
  {C.}~\bibnamefont {Rettinger}}, \bibinfo {author} {\bibfnamefont
  {F.}~\bibnamefont {Schornbaum}}, \bibinfo {author} {\bibfnamefont
  {C.}~\bibnamefont {Schwarzmeier}}, \bibinfo {author} {\bibfnamefont
  {D.}~\bibnamefont {Thönnes}}, \bibinfo {author} {\bibfnamefont
  {H.}~\bibnamefont {Köstler}},\ and\ \bibinfo {author} {\bibfnamefont
  {U.}~\bibnamefont {Rüde}},\ }\bibfield  {title} {\bibinfo {title}
  {{waLBerla: A block-structured high-performance framework for multiphysics
  simulations}},\ }\href
  {https://doi.org/https://doi.org/10.1016/j.camwa.2020.01.007} {\bibfield
  {journal} {\bibinfo  {journal} {Computers \& Mathematics with Applications}\
  }\textbf {\bibinfo {volume} {81}},\ \bibinfo {pages} {478} (\bibinfo {year}
  {2021})},\ \bibinfo {note} {development and Application of Open-source
  Software for Problems with Numerical PDEs}\BibitemShut {NoStop}%
\bibitem [{\citenamefont {Montessori}\ \emph {et~al.}(2024)\citenamefont
  {Montessori}, \citenamefont {La~Rocca}, \citenamefont {Amati}, \citenamefont
  {Lauricella}, \citenamefont {Tiribocchi},\ and\ \citenamefont
  {Succi}}]{montessoriTSLB-pof-2024}%
  \BibitemOpen
  \bibfield  {author} {\bibinfo {author} {\bibfnamefont {A.}~\bibnamefont
  {Montessori}}, \bibinfo {author} {\bibfnamefont {M.}~\bibnamefont
  {La~Rocca}}, \bibinfo {author} {\bibfnamefont {G.}~\bibnamefont {Amati}},
  \bibinfo {author} {\bibfnamefont {M.}~\bibnamefont {Lauricella}}, \bibinfo
  {author} {\bibfnamefont {A.}~\bibnamefont {Tiribocchi}},\ and\ \bibinfo
  {author} {\bibfnamefont {S.}~\bibnamefont {Succi}},\ }\bibfield  {title}
  {\bibinfo {title} {{High-order thread-safe lattice Boltzmann model for high
  performance computing turbulent flow simulations}},\ }\href
  {https://doi.org/10.1063/5.0202155} {\bibfield  {journal} {\bibinfo
  {journal} {Physics of Fluids}\ }\textbf {\bibinfo {volume} {36}},\ \bibinfo
  {pages} {035171} (\bibinfo {year} {2024})}\BibitemShut {NoStop}%
\bibitem [{\citenamefont {Lauricella}\ \emph {et~al.}(2025)\citenamefont
  {Lauricella}, \citenamefont {Tiribocchi}, \citenamefont {Succi},
  \citenamefont {Brandt}, \citenamefont {Mukherjee}, \citenamefont {Rocca},\
  and\ \citenamefont {Montessori}}]{lauricellaTSLB2025arxiv}%
  \BibitemOpen
  \bibfield  {author} {\bibinfo {author} {\bibfnamefont {M.}~\bibnamefont
  {Lauricella}}, \bibinfo {author} {\bibfnamefont {A.}~\bibnamefont
  {Tiribocchi}}, \bibinfo {author} {\bibfnamefont {S.}~\bibnamefont {Succi}},
  \bibinfo {author} {\bibfnamefont {L.}~\bibnamefont {Brandt}}, \bibinfo
  {author} {\bibfnamefont {A.}~\bibnamefont {Mukherjee}}, \bibinfo {author}
  {\bibfnamefont {M.~L.}\ \bibnamefont {Rocca}},\ and\ \bibinfo {author}
  {\bibfnamefont {A.}~\bibnamefont {Montessori}},\ }\href
  {https://arxiv.org/abs/2501.00846} {\bibinfo {title} {{Thread-safe multiphase
  lattice Boltzmann model for droplet and bubble dynamics at high density and
  viscosity contrasts}}} (\bibinfo {year} {2025}),\ \Eprint
  {https://arxiv.org/abs/2501.00846} {arXiv:2501.00846 [physics.flu-dyn]}
  \BibitemShut {NoStop}%
\bibitem [{\citenamefont {Latt}\ \emph {et~al.}(2008)\citenamefont {Latt},
  \citenamefont {Chopard}, \citenamefont {Malaspinas}, \citenamefont
  {Deville},\ and\ \citenamefont {Michler}}]{Latt2008GradBC}%
  \BibitemOpen
  \bibfield  {author} {\bibinfo {author} {\bibfnamefont {J.}~\bibnamefont
  {Latt}}, \bibinfo {author} {\bibfnamefont {B.}~\bibnamefont {Chopard}},
  \bibinfo {author} {\bibfnamefont {O.}~\bibnamefont {Malaspinas}}, \bibinfo
  {author} {\bibfnamefont {M.}~\bibnamefont {Deville}},\ and\ \bibinfo {author}
  {\bibfnamefont {A.}~\bibnamefont {Michler}},\ }\bibfield  {title} {\bibinfo
  {title} {{Straight velocity boundaries in the lattice Boltzmann method}},\
  }\href {https://doi.org/10.1103/PhysRevE.77.056703} {\bibfield  {journal}
  {\bibinfo  {journal} {Phys. Rev. E}\ }\textbf {\bibinfo {volume} {77}},\
  \bibinfo {pages} {056703} (\bibinfo {year} {2008})}\BibitemShut {NoStop}%
\bibitem [{\citenamefont {Dorschner}\ \emph {et~al.}(2015)\citenamefont
  {Dorschner}, \citenamefont {Chikatamarla}, \citenamefont {Bösch},\ and\
  \citenamefont {Karlin}}]{DORSCHNER2015GradBC}%
  \BibitemOpen
  \bibfield  {author} {\bibinfo {author} {\bibfnamefont {B.}~\bibnamefont
  {Dorschner}}, \bibinfo {author} {\bibfnamefont {S.}~\bibnamefont
  {Chikatamarla}}, \bibinfo {author} {\bibfnamefont {F.}~\bibnamefont
  {Bösch}},\ and\ \bibinfo {author} {\bibfnamefont {I.}~\bibnamefont
  {Karlin}},\ }\bibfield  {title} {\bibinfo {title} {{Grad's approximation for
  moving and stationary walls in entropic lattice Boltzmann simulations}},\
  }\href {https://doi.org/https://doi.org/10.1016/j.jcp.2015.04.017} {\bibfield
   {journal} {\bibinfo  {journal} {Journal of Computational Physics}\ }\textbf
  {\bibinfo {volume} {295}},\ \bibinfo {pages} {340} (\bibinfo {year}
  {2015})}\BibitemShut {NoStop}%
\bibitem [{\citenamefont {D'Orazio}\ and\ \citenamefont
  {Succi}(2003)}]{succi2003ThermalBCs}%
  \BibitemOpen
  \bibfield  {author} {\bibinfo {author} {\bibfnamefont {A.}~\bibnamefont
  {D'Orazio}}\ and\ \bibinfo {author} {\bibfnamefont {S.}~\bibnamefont
  {Succi}},\ }\bibfield  {title} {\bibinfo {title} {{Boundary Conditions for
  Thermal Lattice Boltzmann Simulations}},\ }in\ \href@noop {} {\emph {\bibinfo
  {booktitle} {Computational Science --- ICCS 2003}}},\ \bibinfo {editor}
  {edited by\ \bibinfo {editor} {\bibfnamefont {P.~M.~A.}\ \bibnamefont
  {Sloot}}, \bibinfo {editor} {\bibfnamefont {D.}~\bibnamefont {Abramson}},
  \bibinfo {editor} {\bibfnamefont {A.~V.}\ \bibnamefont {Bogdanov}}, \bibinfo
  {editor} {\bibfnamefont {J.~J.}\ \bibnamefont {Dongarra}}, \bibinfo {editor}
  {\bibfnamefont {A.~Y.}\ \bibnamefont {Zomaya}},\ and\ \bibinfo {editor}
  {\bibfnamefont {Y.~E.}\ \bibnamefont {Gorbachev}}}\ (\bibinfo  {publisher}
  {Springer Berlin Heidelberg},\ \bibinfo {address} {Berlin, Heidelberg},\
  \bibinfo {year} {2003})\ pp.\ \bibinfo {pages} {977--986}\BibitemShut
  {NoStop}%
\bibitem [{\citenamefont {Rasin}\ \emph {et~al.}(2005)\citenamefont {Rasin},
  \citenamefont {Miller},\ and\ \citenamefont {Succi}}]{succiPhaseField2005}%
  \BibitemOpen
  \bibfield  {author} {\bibinfo {author} {\bibfnamefont {I.}~\bibnamefont
  {Rasin}}, \bibinfo {author} {\bibfnamefont {W.}~\bibnamefont {Miller}},\ and\
  \bibinfo {author} {\bibfnamefont {S.}~\bibnamefont {Succi}},\ }\bibfield
  {title} {\bibinfo {title} {Phase-field lattice kinetic scheme for the
  numerical simulation of dendritic growth},\ }\href
  {https://doi.org/10.1103/PhysRevE.72.066705} {\bibfield  {journal} {\bibinfo
  {journal} {Phys. Rev. E}\ }\textbf {\bibinfo {volume} {72}},\ \bibinfo
  {pages} {066705} (\bibinfo {year} {2005})}\BibitemShut {NoStop}%
\end{thebibliography}%
\end{document}